\documentclass{jfm}

\usepackage{graphicx}
\usepackage{natbib}
\usepackage{epstopdf}
\AppendGraphicsExtensions{.tif}
\usepackage{amsmath}
\usepackage{flushend}
\usepackage{rotating}

\usepackage{ amssymb, amsfonts}

\usepackage{xcolor}

\ifCUPmtlplainloaded \else
  \checkfont{eurm10}
  \iffontfound
    \IfFileExists{upmath.sty}
      {\typeout{^^JFound AMS Euler Roman fonts on the system,
                   using the 'upmath' package.^^J}%
       \usepackage{upmath}}
      {\typeout{^^JFound AMS Euler Roman fonts on the system, but you
                   dont seem to have the}%
       \typeout{'upmath' package installed. JFM.cls can take advantage
                 of these fonts,^^Jif you use 'upmath' package.^^J}%
      }
  \else
  \fi
\fi

\ifCUPmtlplainloaded \else
  \checkfont{msam10}
  \iffontfound
    \IfFileExists{amssymb.sty}
      {\typeout{^^JFound AMS Symbol fonts on the system, using the
                'amssymb' package.^^J}%
       \usepackage{amssymb}%
       \let\le=\leqslant  
       \let\ge=\geqslant  
      }{}
  \fi
\fi

\ifCUPmtlplainloaded \else
  \IfFileExists{amsbsy.sty}
    {\typeout{^^JFound the 'amsbsy' package on the system, using it.^^J}%
     \usepackage{amsbsy}}
    {\providecommand\boldsymbol[1]{\mbox{\boldmath $##1$}}}
\fi

\newsavebox{\astrutbox}
\sbox{\astrutbox}{\rule[-5pt]{0pt}{20pt}}

\newcommand{\pp}{\ensuremath{\tilde{p}}}

\newcommand{\w}{\ensuremath{\mathcal{W}}}
\newcommand{\dd}{\ensuremath{\mathrm{d}}}
\newcommand{\id}{\ensuremath{\hphantom{.}\mathrm{d}}}
 
 \mathchardef\mhyphen="2D

\newcommand{\solidLine}[1]{\protect\raisebox{0.8ex}{\color{#1}\linethickness{0.5mm}\line(1,0){0.6}}}
\newcommand{\dashLine}[1]{\protect\raisebox{0.8ex}{\color{#1}\linethickness{0.5mm}\line(1,0){0.3}\hspace{0.15cm}\line(1,0){0.3}}}
\newcommand{\dashDotLine}[1]{\protect\raisebox{0.8ex}{\color{#1}\linethickness{0.5mm}\line(1,0){0.3}\hspace{0.15cm}\line(1,0){0.1}\hspace{0.1cm}\linethickness{0.5mm}\line(1,0){0.3}}}
\newcommand{\dotLine}[1]{\protect\raisebox{0.8ex}{\color{#1}\linethickness{0.5mm}\line(1,0){0.1}\hspace{0.1cm}\line(1,0){0.1}\hspace{0.1cm}\line(1,0){0.1}}}

\definecolor{myred}{RGB}{179, 13, 13}
\definecolor{myblue}{RGB}{0, 0, 153}
\definecolor{mygreen}{RGB}{0, 102, 0}
\definecolor{myorag}{RGB}{246, 150, 38}
\definecolor{mypurp}{RGB}{73,0,146}
\definecolor{myyel}{RGB}{235,235,10}
\definecolor{LGrey}{rgb}{.5,.5,.5}

\title[Minimal-channel DNS of spanwise-aligned bars]{
Direct numerical simulation of high aspect ratio spanwise-aligned bars}

\author[M. MacDonald, A. Ooi, R. Garc\'{i}a-Mayoral, N. Hutchins and D. Chung]
{M. MacDonald$^1$\thanks{Email address for correspondence: michael.macdonald@unimelb.edu.au},\ns
A. Ooi$^1$,
R. Garc\'{i}a-Mayoral$^2$,
N. Hutchins$^1$
and D. Chung$^1$}

\affiliation{
$^1$
Department of Mechanical Engineering, University of Melbourne, Victoria 3010, Australia \\
$^2$
Department of Engineering, University of Cambridge, Cambridge CB2 1PZ, UK
}

\pubyear{2015}
\volume{999}
\pagerange{998--999}
\date{?; revised ?; accepted ?}
\begin{document}

\maketitle

%
%
\begin{abstract}
We conduct minimal-channel direct numerical simulations of turbulent flow over two-dimensional rectangular bars aligned in the spanwise direction.
This roughness has been often described as $d$-type, as the roughness function $\Delta U^+$ is thought to depend only on the outer-layer length scale (pipe diameter, channel half height or boundary layer thickness).
This is in contrast to conventional engineering rough surfaces, named $k$-type, for which $\Delta U^+$ depends on the roughness height, $k$.
The minimal-span rough-wall channel  is used to circumvent the high cost of simulating high Reynolds number flows, enabling a range of bars with varying
aspect ratios to be investigated.
The present results show that increasing the trough-to-crest height ($k$) of the roughness while keeping the width between roughness bars, $\w$, fixed in viscous units,
results in non-$k$-type behaviour although this does not necessarily indicate $d$-type behaviour.
  Instead, for deep surfaces with $k/\w\gtrsim 3$, the roughness function appears to depend only on $\w$ in viscous units.
In these situations, the flow no longer has any information about how deep the roughness is and instead can only `see' the width of the fluid gap between the bars.
\end{abstract}

\begin{keywords}

\end{keywords}

%
%
\section{Introduction}
\label{sect:intro}

Turbulent flows bounded by a rough wall are ubiquitous in engineering and geophysical applications. The roughness generally increases the drag force exerted on the wall when compared to a smooth wall, which is often quantified by the (Hama) roughness function, $\Delta U^+$ \citep{Hama54}. This quantity reflects the retardation of the mean streamwise flow over a rough wall compared to a smooth wall, and can be related to the difference in skin-friction coefficients, $C_f$. The superscript $+$ indicates quantities non-dimensionalised on kinematic viscosity $\nu$ and friction velocity $U_\tau\equiv\sqrt{\tau_w/\rho}$, where $\tau_w$ is the wall-shear stress and $\rho$ is the fluid density.
The intuition that increasing the roughness height would increase the drag suggests that the roughness function should scale on some characteristic roughness height $k^+$. In the fully rough regime, in which the skin-friction coefficient no longer depends on the Reynolds number, the roughness function scales as $\Delta U^+=\kappa^{-1}\log(k^+)+B$, where $\kappa\approx0.4$ is the von K{\'a}rm{\'a}n constant and $B$ depends on the rough surface in question \citep{Hama54}. 
If the offset $B$ is known, then extrapolations to engineering roughness Reynolds numbers can be easily performed. Alternatively, the equivalent sand grain roughness $k_s$ can be reported, which relates a given roughness length scale to the sand grain roughness size of \cite{Nikuradse33} as $k/k_s \equiv\exp(-\kappa(3.5+B))$, where the constant 3.5 comes from the difference between the smooth-wall log-law offset ($\approx5$) and Nikuradse's rough-wall constant ($\approx8.5$). These surfaces have been termed $k$-type roughness, due to the dependence of the roughness function on the roughness height.

In contrast to $k$-type roughness, a second type of rough-wall flow was discovered for closely packed rectangular bars aligned in the spanwise direction in pipes \citep{Streeter49,Sams52,Ambrose56}. This was discussed in the seminal work by \cite{Perry69}, who studied these spanwise-aligned bars in a developing turbulent boundary layer. Here, the roughness function was shown to not scale on the roughness height $k^+$, but rather the boundary layer height, $\delta^+$, as $\Delta U^+ = \kappa^{-1}\log(\delta ^+)+B_d$. \cite{Perry69} termed this roughness $d$-type roughness (named after the pipe diameter, due to the earlier pipe flow studies), as it depends on the outer-layer length scale (pipe diameter, boundary layer thickness or channel half-height, $h$).
The flow physics determining how the outer-layer length scale influences $\Delta U^+$  is unclear.  Authors such as \cite{Perry69}, \cite{Cui03} and \cite{Coleman07} suggested that there are stable vortices inside the roughness cavities which are isolated from the flow above. The flow within the roughness canopy (below the roughness crest) would therefore be similar to a lid-driven cavity flow, while the outer-flow would see what is similar to an alternating slip and no-slip boundary condition at the interface. The outer-layer flow would have no information about how deep the cavities are, implying $k$ is not relevant. \cite{Townsend76}, \cite{Djenidi99} and \cite{Jimenez04} proposed some form of ejection of the roughness cavity flow into the outer layer, where these ejections are triggered by an outer-layer-dependent process such as large-scale sweeps, which scales with $d$.

Much of the difficulty in studying $d$-type roughness comes from experimental uncertainty in determining $U_\tau$. An incorrect measure of this quantity directly influences $\Delta U^+\equiv\Delta U/U_\tau$, which can make isolating the effects of $k$ and $h$ problematic. \cite{Jimenez04} reviewed several experimental studies of $d$-type rough surfaces but found that the evidence for the idea that $\Delta U^+$ only scales on $h$ was uncertain. Conventional direct numerical simulations provide an exact estimate of $U_\tau$, but the closely packed nature of the spanwise aligned bars necessitates an extremely dense grid. \cite{Leonardi07} provided one of the first direct numerical simulations (DNS) of a $d$-type surface. These authors were not able to verify if $\Delta U^+$ was a function of the channel half-height $h$, but did show that $\Delta U^+$ was not a function of the roughness height $k^+$. They therefore broadened the definition of $d$-type roughness to be any surface for which $\Delta U^+\ne f(k^+)$.

 The aforementioned expense of the dense grid has made simulating $d$-type roughness unfeasible for many researchers. However, recently it was shown in \cite{Chung15} and \cite{MacDonald17} that a minimal-span channel can be used to directly simulate rough-wall flows, which follows  on from the early work of \cite{Jimenez91} and \cite{Hamilton95} in smooth-wall minimal domains. 
 This technique involves restricting the spanwise domain width, $L_y$, to be much smaller than the channel half height, $h$, with $L_y^+\sim O(100)$ typical. Only the near-wall flow around the roughness is fully resolved while the outer-layer flow is restricted by the narrow domain. As a result, the mean velocity profile of the minimal-span channel deviates from the full-span channel above a vertical (wall-normal) critical height $z_c\approx0.4L_y$. Below $z_c$, the turbulent flow is the same as in a full-span channel, so is regarded as `healthy' turbulence \citep{Flores10}.
 \cite{MacDonald16} successfully used this technique to simulate densely packed sinusoidal surfaces with solidities (frontal roughness area divided by plan area) of up to $\Lambda=0.54$. Given that $d$-type surfaces usually have $\Lambda\ge0.5$ \citep{Jimenez04}, then the minimal-span channel would be highly suitable for such surfaces. As there is explicitly no outer-layer length scale associated with the flow (as there are no outer-layer length scale eddies) in the minimal channel framework, we cannot explicitly test the hypothesis that $\Delta U^+=f(h^+)$. However, by increasing the channel width and therefore the largest captured length scale $z_c$, this can
 serve as a span-independence test to  provide some indication on the influence of the largest length scale in the flow. Moreover, if we consider several rough surfaces with varying $k^+$ then we can examine what functional dependence, if any, $\Delta U^+$ has on $k^+$.

\setlength{\unitlength}{1cm}
\begin{figure}
\centering
	\includegraphics[trim=-25 0 0 0,clip=true]{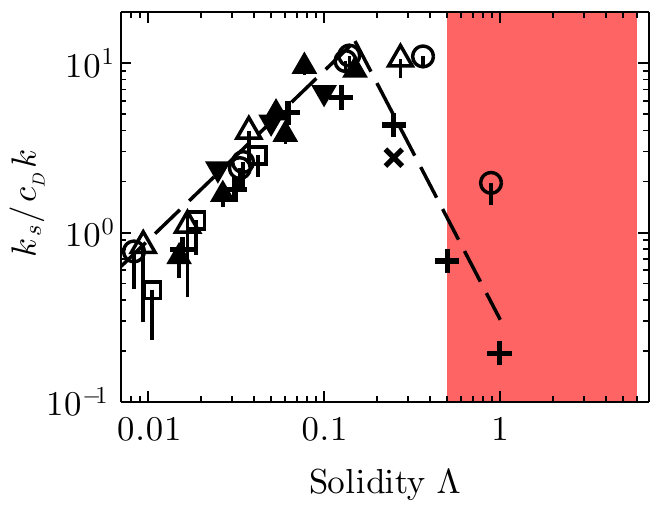}
	\put(-7.2,5){(\emph{a})}
	\\
	\vspace{-0.0\baselineskip}
	\includegraphics[trim=-25 0 0 0,clip=true]{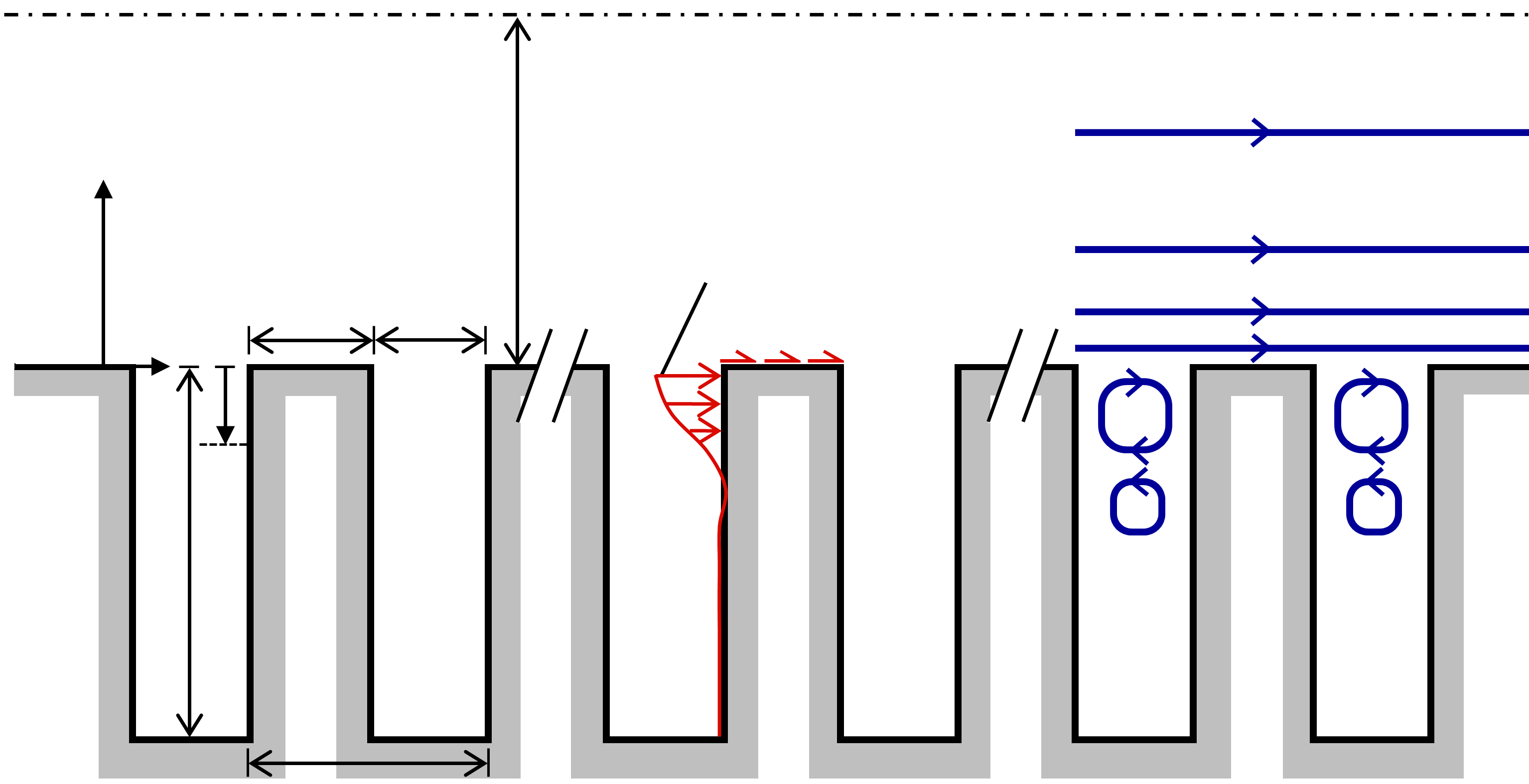}
	\put(-12.55,5.8){(\emph{b})}
	\put(-10.6,4.4){$z$}
	\put(-10.5,3.25){$x$}
	\put(-10.35,1.7){$k$}
	\put(-10.05,2.3){$-\epsilon$}
	\put(-8.8,-0.2){$\lambda$}
	\put(-9.25,3.45){$b$}
	\put(-8.45,3.45){$\w$}
	\put(-7.95,4.4){$h$}
	\put(-12.75,3.35){Smooth-wall}
	\put(-12.75,3.05){origin}
	\put(-6.0,3.35){$\tau_{visc}$}
	\put(-6.15,3.7){$\Delta p$}
	\put(-2.5,5.2){Flow}
	\put(-11.80,3.14){\vector(1,0){0.4}}
	\vspace{-0.5\baselineskip}
\caption{(Colour online) (\emph{a}) Normalised equivalent sand grain roughness against solidity for different rough surfaces, adapted from \cite{Jimenez04}. The solidity of the present rough surfaces ranges over $0.5\le \Lambda\le 6$, shown by the red filled region.
(\emph{b}) Sketch of the present roughness. In all the present roughness simulations, $\w=b=\lambda/2$. Virtual origin in $z$ denoted by $\epsilon$.
Pressure drop across
a single roughness element given by $\Delta p = \Delta P + \Delta \pp$, where $P$ is the mean (driving) pressure component and $\pp$ the fluctuating (periodic) component (see \S\ref{sect:vol}).
 Blue arrows show streamlines commonly used to describe $d$-type roughness.
}
	\label{fig:barSketch}
\end{figure}
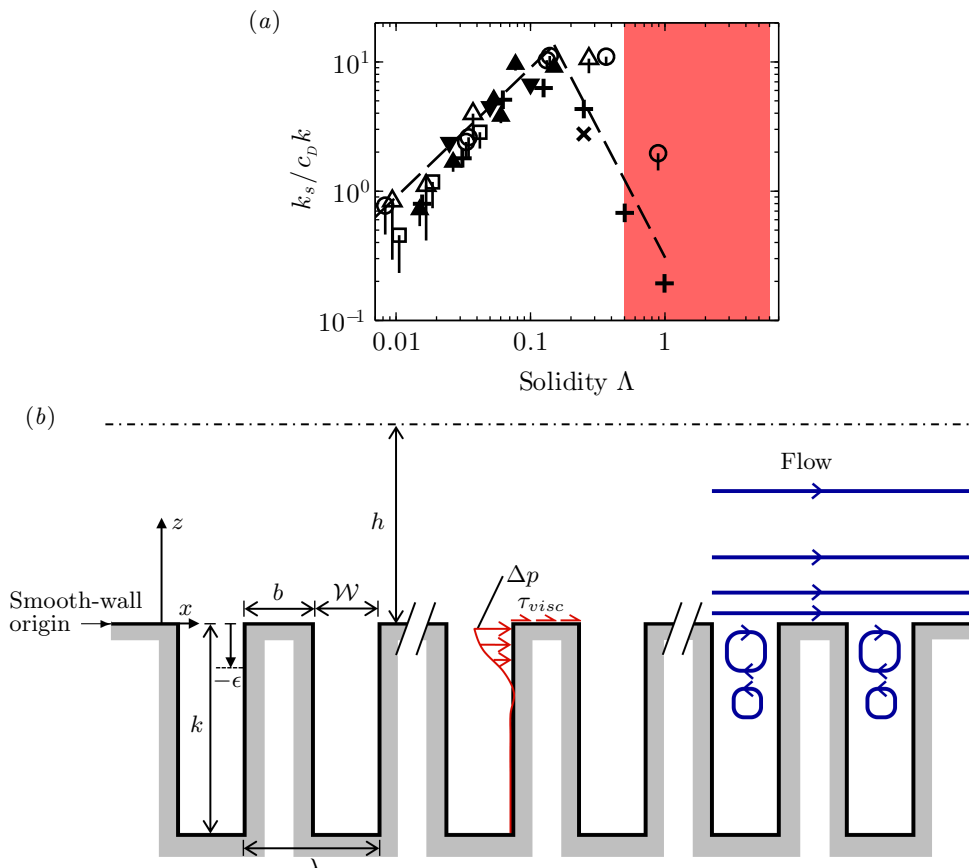

Most previous $d$-type roughness studies use square bars aligned in the spanwise direction, and vary the width of the fluid gap between the bars, $\w$, for a fixed $k$ \citep[e.g.][]{Djenidi99,Cui03,Coleman07,Leonardi07}. The present study varies $k$ for fixed values of $\w$, which corresponds to progressively taller rectangular bars.  The aspect ratio, or solidity, $\Lambda=k/(2\w)$, of the bars studied herein is exceptionally large, with values of 0.5 up to 6. Figure \ref{fig:barSketch}(\emph{a}) is adapted from figure 1 of \cite{Jimenez04}, and shows the equivalent sand grain roughness plotted against solidity for different rough surfaces. The drag coefficient $C_D$ is present to account for different roughness geometries. In the so-called sparse regime ($\Lambda\le0.15$), the equivalent sand grain roughness is seen to scale with $\Lambda$ for several different surfaces, or in other words the drag increases with increasing roughness frontal area. The so-called dense regime ($\Lambda\ge0.15$) is less studied owing to the high experimental costs and difficulties mentioned above, and there is no known scaling argument in this region \citep{Jimenez04}. The dashed line in figure \ref{fig:barSketch}(\emph{a}) in the dense regime shows a $\Lambda^{-2}$ scaling, although different powers have been proposed \citep{Jimenez04}.
 The solidity of the rough surfaces in the present study is indicated by the red filled region, and can be seen to have much larger aspect ratios than previous studies. 
 The sketch in figure \ref{fig:barSketch}(\emph{b}) of the present roughness shows the flow patterns often  used to describe $d$-type roughness \citep{Jimenez04}. Intuitively, the vortices inside the roughness cavities would likely scale on the cavity width $\w$ which could suggest some kind of $\Delta U^+=f(\w^+)$ dependence. Rather than create a new `$\w$-type' roughness classification, we will follow \cite{Leonardi07} and continue to use the term $d$-type roughness, to refer to a surface where $\Delta U^+\ne f(k^+)$.

As mentioned earlier, several authors have made comparisons between the roughness cavity flow in $d$-type surfaces and lid-driven cavity flows  \citep{Perry69,Cui03,Coleman07}. While most lid-driven cavity flows consider square cavities, some consider varying aspect ratios of $k/\w$ \citep[e.g.][]{Cheng06,Patil06}. These deep cavities are observed to have a cascade of alternately rotating vortices that extend down to the lower wall. Importantly, \cite{Cheng06} showed that the number of vortices present is approximately equal to the aspect ratio $k/\w$ which supports the present view that the flow could somehow be dependent on $\w$. The strength of these vortices diminishes rapidly with depth however, to the point that below a few cavity widths they have near-zero velocity. If this is the case with the present bar roughness, then increasing $k$ for a fixed $\w$, such that $k/\w$ exceeds approximately 2, would have no impact on the flow.

While previous $d$-type studies have all used two-dimensional spanwise bars, tightly packed three-dimensional roughness has been suggested to exhibit certain $d$-type characteristics, notably the skimming behaviour of the flow over the roughness \citep{Hunter92,Yang16,Sadique17}.
In particular, \cite{Sadique17} conducted large-eddy simulations (LES) of flow over high aspect ratio (large $k/\w$) three-dimensional rectangular prisms, showing that the velocity variations within the roughness canopy were confined to a region of  distance $\w$ from the roughness crest. The authors suggested that below this point the roughness canopy is sheltered or inactive and does not contribute to drag production, so that $k$ is not relevant.  The aim of the present study is therefore to vary $k$ for fixed $\w$ with the aspect ratio $k/\w$ being large, enabling a systematic investigation into these two roughness length scales.

%
%
\section{Numerical procedure}
\label{sect:numerics}
The numerical method used in this study is described and validated in \cite{Chan15} and \cite{MacDonald16}
 This is a second-order finite volume code \citep{Ham04,Mahesh04} 
which directly solves the Navier--Stokes equations. A half-height (open) channel is used whereby a slip wall is positioned at $z=h$,  while a no-slip impermeable wall is used for the spanwise-aligned bars with the trough at $z=-k$ and crest at $z=0$. Periodic boundary conditions are applied in the streamwise and spanwise directions. 
 The flow is driven by a prescribed constant mass flux, so that the driving pressure gradient of the channel, $G_x(t) = -\dd P/\dd x$, varies at each time step.
This mass flux is set via trial and error such that the friction Reynolds number, $Re_\tau=U_\tau h/\nu\approx395$ for all cases.
This Reynolds number was selected as it was shown in \cite{Chan15} that the roughness function, $\Delta U^+$, is overestimated for $Re_\tau=180$ simulations, becoming friction Reynolds number invariant for $Re_\tau\gtrsim360$
 for roughness with matched viscous dimensions (i.e. the same  $k^+$ and $\w^+$).
Note that the primary cause for the overestimate in $\Delta U^+$ is actually due to an overshoot of the logarithmic region of low $Re_\tau$ smooth-wall flows, as opposed to the rough-wall flow. Therefore, while this observation in \cite{Chan15} that $\Delta U^+$ becomes invariant for $Re_\tau\gtrsim360$ comes from sinusoidal roughness simulations, it should still apply to the present spanwise bar roughness as it is predominantly due to a smooth-wall flow effect.
Table \ref{tab:sims} details the simulations that were performed in this study. The expected full-span bulk velocity, $U_{bf}^+=\int U_{f}^+\mathrm{d}z^+/h^+$, is given, where the expected full-span velocity profile $U_f$ is defined such that the simulation data from the minimal channel is used for $z<z_c$, while the composite velocity profile of \cite{Nagib08} for full-span channel flow is used  for $z>z_c$ where the log-law offset constant is set such that $U_f$ is continuous at $z=z_c$.

\begin{table}
\centering
\begin{tabular}{c c c c c c c c c c c c c c c}
$\w^+$	&	$k^+$	& $\lambda^+$  & $\Lambda$	& $\frac{h+k}{k}$	& $L_x^+$	& $L_y^+$	& $N_x$	& $N_y$	& $N_z$	& $N_{zk}$ 	& $\Delta x^+$	& $\Delta z_h^+$  & $U_{bf}^+$ & $\Delta U^+$\\[-0.2em]
 \hline
- 	& - 	& - 	& 0	& -	& 1000	& 153	& 1600	& 34		& 120	& - 	& 0.63	& 12.9	& 17.6	& - \\
10	& 5 	& 20 	& 0.25& 80.0	& 1000	& 153	& 1600	& 34		& 120	& 12 	& 0.63	& 12.9	& 16.7	& 0.9 \\
10	& 10 	& 20 	& 0.5	& 40.5	& 1000	& 153	& 1600	& 34		& 120	& 24 	& 0.63	& 12.9	& 16.1	& 1.5 \\
10	& 20 	& 20 	& 1.0	& 20.8	& 1000	& 153	& 1600	& 34		& 120	& 48 	& 0.63	& 12.9	& 15.6	& 2.0 \\
10	& 30 	& 20 	& 1.5	& 14.2	& 1000	& 153	& 1600	& 34		& 120	& 72 	& 0.63	& 12.9	& 15.2	& 2.5 \\
10	& 40 	& 20 	& 2.0	& 10.9	& 1000	& 153	& 1600	& 34		& 120	& 96 	& 0.63	& 12.9	& 15.1	& 2.6 \\
10	& 60 	& 20 	& 3.0	& 7.58	& 1000	& 153	& 1600	& 34		& 120	& 144 	& 0.63	& 12.9	& 15.1	& 2.6 \\
20	& 60 	& 40 	& 1.5	& 7.58	& 1000	& 153	& 800		& 34		& 120	& 144 	& 1.25	& 12.9	& 15.0	& 2.7 \\
20	& 120 	& 40 	& 3.0	& 4.29	& 1000	& 153	& 800		& 34		& 120	& 195 	& 1.25	& 12.9	& 14.8	& 2.8 \\
-	& - 	& - 	& 0	& -	& 1000	& 306	& 320		& 68		& 120	& - 	& 3.13	& 12.9	& 17.7	& - \\
50	& 150 	& 100 	& 1.5	& 3.63	& 1000	& 306	& 320		& 68		& 120	& 360 	& 3.13	& 12.9	& 14.4	& 3.3 \\
50	& 300 	& 100 	& 3.0	& 2.32	& 1000	& 306	& 320		& 68		& 120	& 720 	& 3.13	& 12.9	& 14.4	& 3.2 \\
100	& 50 	& 200 	& 0.25& 8.90	& 1000	& 306	& 320		& 68		& 120	& 112 	& 3.13	& 12.9	& 13.6	& 4.1 \\
100	& 150 	& 200 	& 0.75& 3.63	& 1000	& 306	& 320		& 68		& 120	& 229 	& 3.13	& 12.9	& 12.9	& 4.0 \\
100	& 300 	& 200 	& 1.5	& 2.32	& 1000	& 306	& 320		& 68		& 120	& 302 	& 3.13	& 12.9	& 12.6	& 3.9 \\
100	& 600 	& 200 	& 3.0	& 1.66	& 1000	& 306	& 320		& 68		& 120	& 452 	& 3.13	& 12.9	& 12.6	& 3.9 \\
100	& 1200& 200 & 6.0	& 1.33	& 1000	& 306	& 320		& 68		& 120	& 752 	& 3.13	& 12.9	& 12.5	& 3.9 \\
-	& - 	& - 	& 0	& -	& 1600	& 612	& 384		& 136		& 162	& - 	& 4.17	& 7.5		& 17.6	& - \\
100	& 300 	& 200 	& 1.5	& 2.32	& 1600	& 612	& 512		& 136		& 162	& 214 	& 3.13	& 7.5		& 12.5	& 3.8 \\
200	& 300 	& 400 	& 0.75& 2.32	& 1600	& 612	& 384		& 136		& 162	& 214 	& 4.17	& 7.5		& 9.32	& 5.4 \\
200	& 600 	& 400 	& 1.5	& 1.66	& 1600	& 612	& 384		& 136		& 162	& 334 	& 4.17	& 7.5		& 9.49	& 5.5 \\
200	& 1200& 400	& 3.0	& 1.33	& 1600	& 612	& 384		& 136		& 162	& 574 	& 4.17	& 7.5		& 9.53	& 5.5 \\
\end{tabular}
\vspace{-0.3\baselineskip}
\caption{Description of the different simulations performed. 
Refer to figure \ref{fig:barSketch} for roughness description. Other symbols:
$\Lambda=k/\lambda$, solidity;
$N_x$, $N_y$, $N_z$, number of cells in streamwise, spanwise and vertical (wall-normal) direction (above roughness crest);
$N_{zk}$, number of cells in the vertical direction below the roughness crest;
$\Delta x^+$ is the streamwise grid spacing
and
$\Delta z_h^+$ is the vertical grid spacing at the channel centre. 
The spanwise grid spacing is always $\Delta y^+=4.5$ while
the vertical grid spacing at the roughness crests is always $\Delta z_w^+=0.4$.
$U_{bf}^+$ is the expected full-span bulk velocity using a composite velocity profile
and $\Delta U^+$ is the roughness function computed from the difference in smooth- and rough-wall velocities evaluated at $z_c^+$.
All simulations conducted at $Re_\tau\approx395$.
}
\label{tab:sims}
\end{table}

A cell-to-cell expansion ratio of approximately 1.028 is used in the vertical (wall-normal) direction above the roughness crest, resulting in a fairly large grid spacing at the channel centreline. However, the grid spacings below $z_c$ are  such that $\Delta z^+$ only increases beyond conventional DNS spacings above the vertical critical height, $z_c$. As the region  of the flow above $z_c$ is already altered due to the nature of the minimal channel, these spacings should have negligible impact on the near-wall flow of interest.
 The verticall mesh spacing below the roughness crest is approximately constant at $\Delta z_w^+$ for cases with $k^+<120$. Cases with larger $k^+$ have a larger vertical spacing towards the centre of the roughness cavity, with $\Delta z^+\approx 2$, however at the roughness crest and trough it is $\Delta z^+\approx0.4$.

The minimal-span channel is used so that the streamwise and spanwise domain sizes are relatively small compared to conventional (full-span) channels. The recommendation in \cite{Chung15} is typically used to determine the spanwise domain width for $k$-type roughness, namely $L_y\gtrsim \max(100\nu/U_\tau,k/0.4,\lambda_{r,y}$), where $\lambda_{r,y}$ is a characteristic roughness spanwise length scale. For the present two-dimensional roughness, $\lambda_{r,y}\rightarrow\infty$ and it is likely that another length scale would take precedence, anticipated to be $\w$ for $\w\ll k$. This $\lambda_{r,y}$ constraint will therefore be ignored.
The second constraint comes from ensuring the roughness is submerged in healthy turbulence i.e. $k< z_c=0.4L_y$. Given that this is likely more applicable for $k$-type roughness and the present roughness may be more dependent on $\w$, we will instead ensure $L_y>O(\w)$, where the independence of the flow below $z_c$ that sets $\Delta U^+$ is more closely examined in the next section.
 The streamwise length should satisfy $L_x\gtrsim\max(3L_y,1000\nu/U_\tau,\lambda_{r,x})$ \citep{MacDonald17}, where for the present roughness with relatively narrow streamwise wavelengths of $\lambda_{r,x}^+\le 200$, the second constraint is the limiting one. 
 A series of larger bars with $\lambda_{r,x}=400$ have also been studied, where the spanwise channel width of $L_y^+=612$ used for these cases necessitates a longer streamwise domain length of $L_x^+=1600$.
  Smooth-wall channel simulations with matched domain sizes have also been conducted, to ensure that the differences between the smooth- and rough-wall flows are due to the roughness alone and not the channel span.

\subsection{Effect of channel width}
\label{ssect:width}

\setlength{\unitlength}{1cm}
\begin{figure}
\centering
	\includegraphics[width=0.49\textwidth]{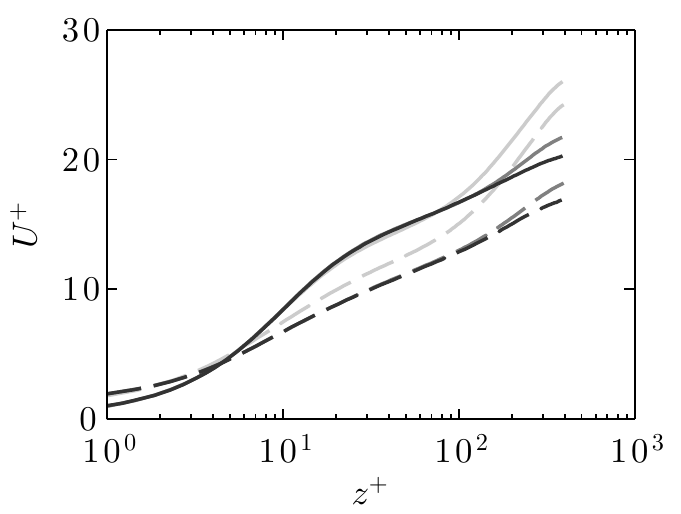}
	\includegraphics[width=0.49\textwidth]{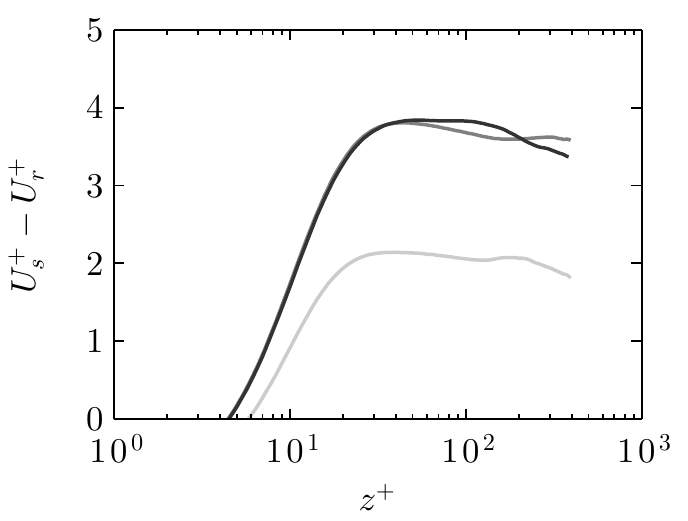}
	\put(-13.6,4.75){(\emph{a})}
	\put(-9.1,2.8){\vector(1,0){1.2}}
	\put(-8.5,2.5){{\scriptsize{$L_y^+=153$}}}
	\put(-8.2,2.25){{\scriptsize{$306,612$}}}
	\put(-6.5,4.75){(\emph{b})}
	\put(-2.7,2.2){{\scriptsize{$L_y^+=153$}}}
	\put(-3.1,4.0){{\scriptsize{$L_y^+=306,612$}}}
	\vspace{-0.5\baselineskip}
\caption{(\emph{a}) Mean velocity profile for smooth-wall (solid) and rough-wall (dashed) minimal channels  with $\w^+\approx100$, $k^+ \approx 300$.
Darker grey refers to increasing channel width (table \ref{tab:sims}).
(\emph{b}) Difference in smooth-wall and rough-wall velocity. The origin in $z^+$ is at the roughness crest.
}
	\label{fig:velWidth}
\end{figure}
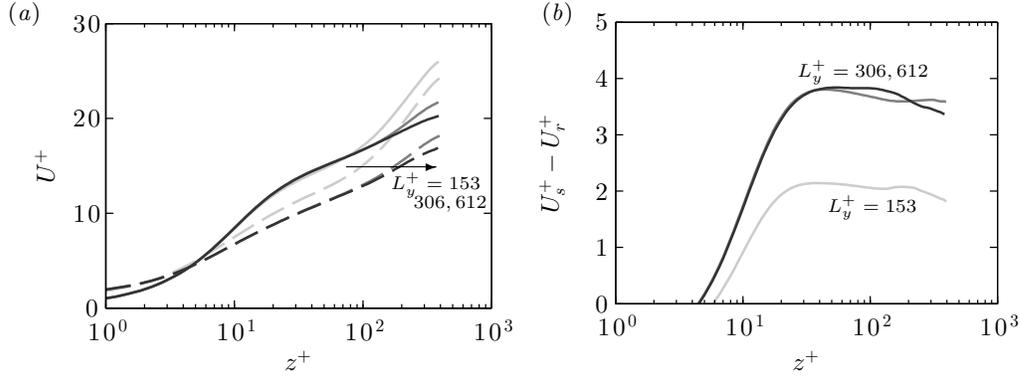

The flow of the minimal-span channel is intrinsically related to the spanwise width, with the critical height scaling as $z_c=0.4L_y$. Previous studies support the view 
that the roughness crest must be submerged in healthy turbulence, requiring $z_c^+>k^+\Rightarrow L_y^+>k^+/0.4$ \citep{Chung15}.
This condition ensures that the roughness sublayer (the region of flow directly affected by the roughness) is fully captured,
the height of which scales with the roughness height for conventional $k$-type roughness \citep{Raupach91,Flack07}.
For the present roughness however, the flow in the troughs would become essentially stationary as the bars become increasingly tall.
This means that the roughness sublayer would be a region of altered turbulence close to the bar crests, with a region of quiescent flow below this.
The roughness sublayer would instead likely scale on the spacing between the bars, $\w$, with $k$ no longer altering the dynamics of the flow.
We will therefore re-examine the relationship needed to capture the roughness sublayer for the present deep bars.

Figure \ref{fig:velWidth}(\emph{a}) shows the effect of increasing channel width on the mean velocity profile for bar roughness with $\w^+\approx 100$, $k^+ \approx 300$. As observed in previous minimal-span channels, increasing the channel width $L_y^+$, and hence critical height $z_c$, reduces the  centreline velocity in the altered outer layer. A larger proportion of energy containing eddies is captured by the wider channels and the results tend to match more closely with conventional full-span channel flow. The rough-wall flow with the smallest width of $L_y^+=153\approx1.5\w^+$ (light grey dashed line)  is seen to result in a slightly increased mean velocity below the critical height of $z_c^+\approx0.4L_y^+\approx61$, when compared to the larger width cases. This is not observed in the smooth-wall flow (solid lines) so that the velocity difference between smooth- and rough-wall flows is nearly $2U_\tau$ lower for $L_y^+=153$ (figure \ref{fig:velWidth}\emph{b}). When $L_y^+\ge306\approx3\w^+$, little difference is seen with increasing channel width.
That is, we have obtained a $\Delta U^+$ that is insensitive to the minimal span and the roughness sublayer is completely captured.
In our simulations, scales larger than $L_y$ are, from the point of view of the roughness, much larger and perceived as effectively infinitely long.
In terms of the critical height, $z_c$, the above results suggest that for spanwise bar roughness we require $z_c\gtrsim 1.2 \w$, or that the height of the healthy turbulence region above the bar crests must be greater than the fluid width between bars, $\w$. 
These results can also be interpreted as indicating that a large scale separation between the largest turbulent length scale, $z_c$, and the roughness length scale, $\w$, is not necessary to ensure that the roughness effects are fully captured.

The classical description of $d$-type roughness is that the roughness function scales on the largest turbulent length scale, which is conventionally the channel half height, $h$. In minimal-span channels the critical height $z_c$ can be regarded as the largest captured turbulent length scale in the flow as opposed to $h$. 
 The above results, however, show that increasing $z_c$ above $1.2\w$ has diminishing influence on the flow. 
 This suggests that the largest turbulent length scales of the flow are independent of the present bar roughness geometry.

%
%
\section{Drag, mean pressure gradient and effective channel height}
\label{sect:vol}

The flow in periodic channels is driven by a mean pressure gradient
$-\dd P/\dd x$, required to balance the friction and pressure drag on the walls.
Numerical simulations are typically run by imposing a uniform body force $G_x(t)=-\dd P/\dd x$ to the entire fluid domain. In the
present simulations, $G_x(t)$ fluctuates in time in order to obtain a constant flow rate through the domain,
however we will consider the time-averaged mean $\overline{G_x}$.
If $D$ is the total drag force on a wall over the periodic box length $L_x$ and width $L_y$, then the drag
per unit plan area, $\tau_w=D/L_x L_y$, can be calculated by an integral force
balance on a periodic volume of the channel.
That is, for an open (half-height channel), the drag force $D$ must equal the pressure difference applied to the frontal (spanwise--vertical) fluid areas of the front and back of the periodic domain, $-\overline{G_x} L_x A$,
and we obtain
\begin{eqnarray}
\tau_w=\frac{D}{L_x L_y}=-\frac{\overline{G_x} A}{L_y},
\label{eqn:tauw}
\end{eqnarray}
The value of both $D$ and $\tau_w$ therefore depends
on the frontal area $A$ of the cross section chosen to delimit the periodic domain or,
alternatively, the mean half-height in the chosen section, $A/L_y$.

\setlength{\unitlength}{1cm}
\begin{figure}
\centering
	\includegraphics{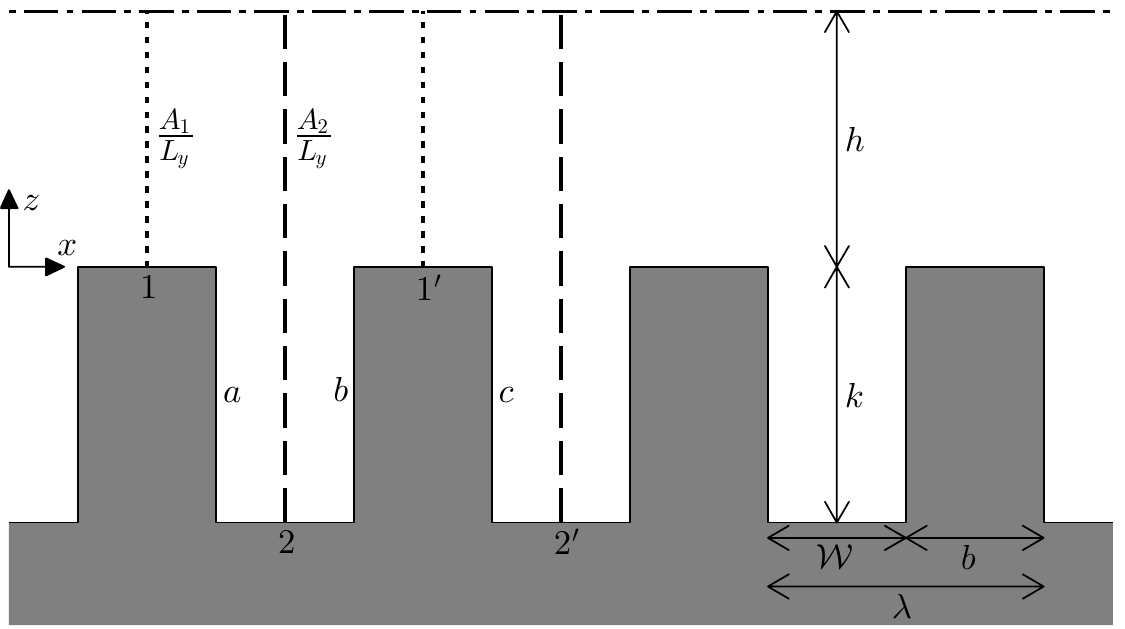}
	\vspace{-0.5\baselineskip}
\caption{Sketch showing the two different choices of the periodic domain, either from the roughness crests, 1--1', or from the roughness troughs, 2--2'.
$a$, $b$ and $c$ indicate the vertical surfaces of the roughness on which the pressure drag acts.
}
	\label{fig:drag_sketch}
\end{figure}

 For the present spanwise bars, the value of
$\tau_w$ depends on whether the delimiting cross section is set at a streamwise location
corresponding to roughness crests, as in section 1--1' in figure \ref{fig:drag_sketch}, yielding
\begin{eqnarray}
\tau_{w,1}=-\frac{\overline{G_x} A_1}{L_y}=-\overline{G_x} h,
\label{eqn:tau1}
\end{eqnarray}
or to roughness troughs, as in section 2--2' in figure \ref{fig:drag_sketch}, yielding
\begin{eqnarray}
\tau_{w,2}=-\frac{\overline{G_x} A_2}{L_y}=-\overline{G_x} (h+k).
\label{eqn:tau2}
\end{eqnarray}
Note that, while $G_x$ is imposed during the simulation, $\tau_w$ is merely obtained \emph{a
posteriori} during post-processing.

The difference between $\tau_{w,1}$ and $\tau_{w,2}$ can also be observed if the drag is obtained from the integral of the
pressure and shear stresses at the wall. For clarity, the true (total) pressure, $p(x,y,z,t)$, can be decomposed into two components; the driving (mean) pressure, $P(x,t)$ and the fluctuating (periodic) pressure, $\pp(x,y,z,t)$. In most DNS codes, including the one used in this study and our previous works \citep{Chan15,MacDonald16}, the driving pressure $P$ is an input into the simulation through the uniform body force $G_x(t) = -\dd P/\dd x$ in the fluid volume. The pressure that is solved via the Poisson solver in the code at each time step is therefore only the periodic component, $\pp$.
In the example sketched in figure
\ref{fig:drag_sketch}, if we set the domain as between sections 1 and 1', we obtain a
drag $D_1$ of
\begin{eqnarray}
D_1=\tau_v L_x L_y - p_{a} k L_y + p_{b} k L_y,
\label{eqn:D1}
\end{eqnarray}
where $\tau_v$ is the $xyt$-mean longitudinal viscous shear stress and $p_a$ and $p_b$ are the $yzt$-mean total pressures acting on the trailing and leading edges of the roughness, respectively. We can decompose the pressure into the driving and periodic components,
\begin{eqnarray}
D_1&=\tau_v L_x L_y + (P_b-P_a+\pp_b-\pp_a) k L_y,
\label{eqn:D1_decomp}
\end{eqnarray}
where $P_b-P_a=\overline{G_x}\w$.
In contrast, if we set the domain as between sections 2 and 2', we obtain
a drag $D_2$ of
\begin{eqnarray}
D_2 &=\tau_v L_x L_y + (P_b-P_c+\pp_b-\pp_c)k L_y.
\label{eqn:D2}
\end{eqnarray}
Due to the periodic nature of the channel, the fluctuating pressures on the back faces of the roughness will be equal, $\pp_c=\pp_a$. However, the difference in driving pressures will be negative, $P_b-P_c=-\overline{G_x}(\lambda-\w)$, or $P_b-P_c=-(P_b-P_a)$ when $\w=\lambda/2$. This occurs because the leading face of the roughness, $b$, is now upstream of the back face, $c$ for section 2--2'.
This means that
\begin{eqnarray}
D_2 &=& \tau_v L_x L_y + (-(P_b-P_a)+\pp_b-\pp_a)k L_y\\
&=& D_1 -\overline{G_x} kL_xL_y.
\label{eqn:D1_D2}
\end{eqnarray}
Hence, the drag per unit plan area, $\tau_w=D/L_xL_y$, for this section is $\tau_{w,2}=\tau_{w,1}+ \overline{G_x} \, k$, which
agrees with the difference observed between (\ref{eqn:tau1}) and (\ref{eqn:tau2}).
In sum, while the mean pressure gradient is unequivocally  defined in a periodic channel, the
measured drag depends on the periodic box chosen as the domain of study,
relative to the non-homogeneous features along the flow direction. This may seem puzzling, as the drag is after all the force exerted on the wall, and it would appear
contradictory that it depended on whether it is measured, say, between sections 1 and 1'
or 2 and 2' in figure \ref{fig:drag_sketch}. 
Note that, for a given channel, the difference
in $\tau_w$ obtained from choosing different delimiting sections remains the same and does
not vanish when choosing very long domains spanning many periodic boxes. 
This ambiguity does not occur in zero-pressure-gradient boundary layers or other flows
not driven by a mean pressure gradient, such as Couette flows. In a rough-wall Couette flow, the total pressures
acting on periodically repeated features, such as $p_\mathrm{a}$ and $p_\mathrm{c}$ in
figure \ref{fig:drag_sketch}, would be identical as the driving pressure component is zero.
As such, (\ref{eqn:D1_decomp}) and (\ref{eqn:D2}) would agree as we would only  be left
with the periodic component of pressure, $\pp_a=\pp_c$. 

Both (\ref{eqn:tau1}) and (\ref{eqn:tau2}) are also  
compatible with the volume integration of the streamwise momentum equation with imposed uniform body force $\overline{G_x}$.
This yields $\widetilde{D}=V_f \overline{G_x} \neq D_1\neq D_2$, where 
$V_f$ is the volume of fluid, uniquely defined regardless of whether the fluid section is chosen as 1--1' or 2--2'.
$\widetilde{D}$ is the force experienced by the fluid
that consists of only the periodic pressure and viscous forces.
This is in contrast to the total drag force exerted on the wall,  $D_1$ and $D_2$ in (\ref{eqn:D1_decomp}) and (\ref{eqn:D2}) respectively, which also includes the driving pressure force.

Equation (\ref{eqn:tauw}) reveals that the relation between the
mean pressure gradient $\overline{G_x}$ and the mean wall stress $\tau_w$ is unequivocal only if the
cross-sectional area of the channel remains constant along $x$, as mentioned by \cite{Saito12}.
 This was the case of the
sinusoidal roughness of \cite{Chan15} and \cite{MacDonald16}, and is also approximately
the case of many rough three-dimensional surfaces, for which the longitudinal change of
the cross-sectional area remains relatively small. In this case, $\widetilde{D}=V_f \overline{G_x} = D_1 = D_2$, and
$V_f = AL_x$ uniquely.  The effect has therefore often been
overlooked in the literature, but it is important for the present spanwise bars, especially
when $k$ is comparable to $h$.

The only unambiguous measure of the drag in our channels
is therefore $\overline{G_x}$, but a definition for $\tau_w$ is needed for any scaling of flow variables
in inner units. In \cite{Chan15} and \cite{MacDonald16}, $D$ and $\tau_w$ could be defined
unambiguously based on the mean channel height or hydraulic radius based on a uniform $A$, which was in that case $h+k/2$. This also
enabled the value of $U_\tau$ to be established naturally as well as a virtual origin $z=0$ at that height.
 The same is not possible for the present spanwise bars. However, these configurations
are densely packed, and we will see in \S\ref{sect:results} that very little
flow penetrates into the cavities. These conditions suggest that, from the perspective of
the turbulent flow, it is more sensible to picture the pressure gradient as driving the flow
through the sections above the bars, such as 1--1' in figure \ref{fig:drag_sketch}.
We therefore choose to
define $z=0$ at the bar tops, which then defines $U_\tau$ from (\ref{eqn:tau1}).

\setlength{\unitlength}{1cm}
\begin{figure}
\centering
	\includegraphics[trim=-7 -4 0 0, clip =true,width=\textwidth]{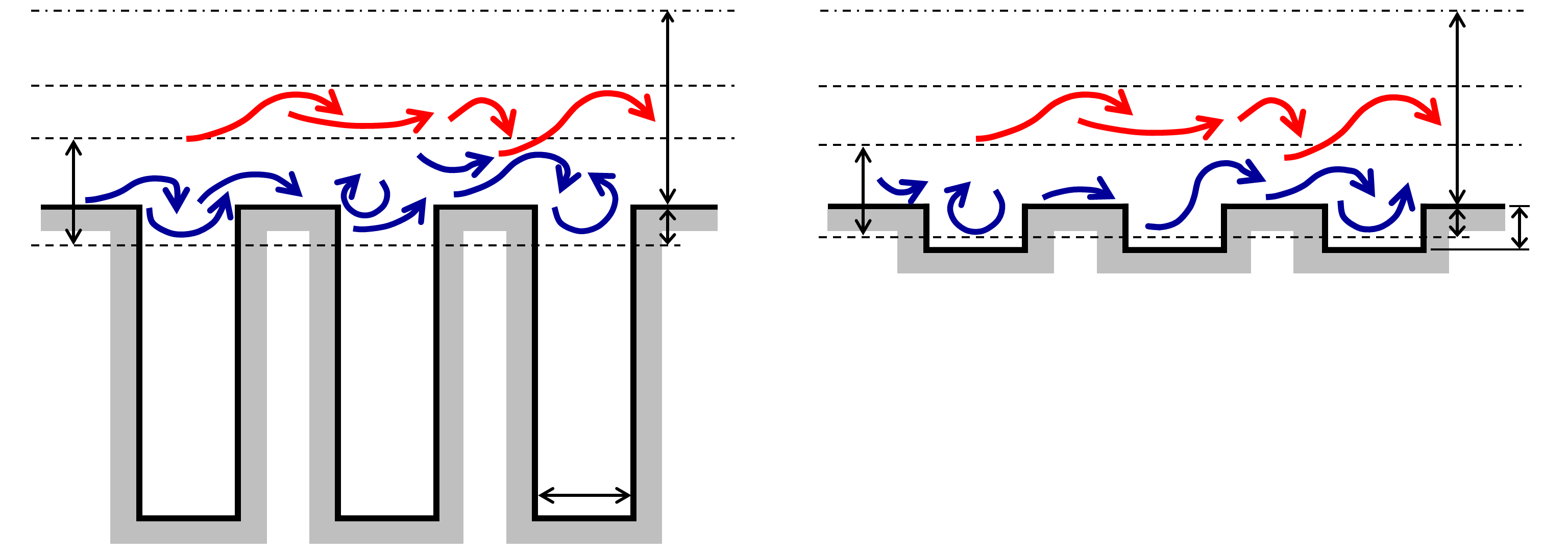}
	\put(-13.63,4.6){(\emph{a})}
	\put(-12.63,3.38){\fontsize{8}{7.2}\selectfont Roughness sublayer}
	\put(-12.15,2.4){\footnotesize Quiescent }
	\put(-12.1,2.05){\footnotesize flow}
	\put(-12.85,3.78){\footnotesize Log layer}	
	\put(-13.52,3.2){$\sim$$3k_s$}
	\put(-8.55,0.65){$\w$}
	\put(-7.6,3.8){$h$}
	\put(-7.57,2.8){$\epsilon$}
	\put(-6.8,4.6){(\emph{b})}	
	\put(-5.85,3.33){\fontsize{8}{7.2}\selectfont Roughness sublayer}	
	\put(-6.10,3.75){\footnotesize Log layer}
	\put(-6.72,3.2){$\sim$$3k_s$}	
	\put(-0.3,2.77){$k$}
	\put(-0.82,2.83){$\epsilon$}
	\put(-0.87,3.75){$h$}
	\vspace{-0.5\baselineskip}
\caption{(Colour online) Sketch showing the roughness sublayer for (\emph{a}) the present deep roughness, where $k_s\propto\w$ (\S\ref{ssect:du}), and (\emph{b}) conventional $k$-type roughness, where $k_s\propto k$.
}
	\label{fig:roughSublayer}
\end{figure}
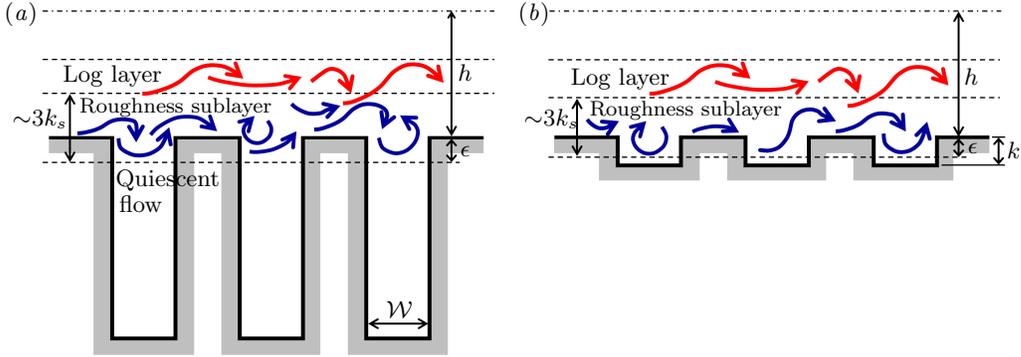

The blockage ratio is typically defined using the roughness height, with the largest blockage for the present cases being approximately $k/(h+k)\approx1/1.33$. This is exceptionally large, and is far outside the recommendation made in \cite{Jimenez04} of $k/(h+k)\lesssim1/40$.
 However, this is a simple geometric measure that  is used to represent the scale separation between the roughness sublayer (a dynamic property) and the channel half height. This scale separation is necessary to ensure that outer-layer similarity is realised and that standard wall-turbulence features like the log layer are observed. The roughness sublayer scales with the equivalent sand grain roughness, $k_s$, which in turn scales with the roughness height $k$ for conventional $k$-type roughnesses  \citep{Raupach91,Flack07}. If the roughness sublayer (of assumed size $3k_s$ or $3k$) is below the mid point of the logarithmic layer ($\tfrac{1}{2}0.15h$), then this leads to the often quoted $k/h\lesssim1/40$ ratio \citep[p.~175]{Jimenez04}. For the present deep spanwise bars, it will be shown that the equivalent sand grain roughness (and hence the approximate thickness of the roughness sublayer) scales as $k_s\approx 0.21\w$ and that the flow only depends on the distance between the bars, $\w$. The roughness sublayer is therefore a small region close to the bar crests, with a region of quiescent flow below this (figure \ref{fig:roughSublayer}). Ensuring sufficient scale separation (or minimal `blockage') for this flow using the same approach as before ($3k_s\approx0.63\w \lesssim \tfrac{1}{2}0.15h$)  therefore leads to a recommendation of $\w/h\lesssim 1/8.4$. The largest value for $\w^+\approx200$ is $\w/h\approx1/2$, which is  now closer to this recommendation. Moreover, numerical studies of internal flows with conventional $k$-type roughness often use larger blockage ratios than the 1/40 recommendation, with 
 1/10 in \cite{Mayoral11jfm}, 
 1/8 in \cite{Leonardi10} and 1/6.75 in \cite{Chan15}. These studies still observe outer-layer similarity, suggesting that the constrained nature of internal flows can support larger blockage ratios compared to external (boundary layer) flows.
 Finally, \S\ref{ssect:width} shows that increasing the channel spanwise width, and hence the largest turbulent length scale in the minimal channel, beyond $z_c\approx 0.4L_y\gtrsim1.2\w$ does not result in significant changes to the near-wall flow. As already discussed, this therefore indicates that we have achieved sufficient scale separation. Ultimately, the large $k/h$ geometric blockage ratios do not appropriately characterise the dynamics of the flow for the present deep bars.

%
%
\section{Bar roughness results}
\label{sect:results}

\subsection{Parameter space}
\label{ssect:paramSpace}

\setlength{\unitlength}{1cm}
\begin{figure}
\centering
	\includegraphics[]{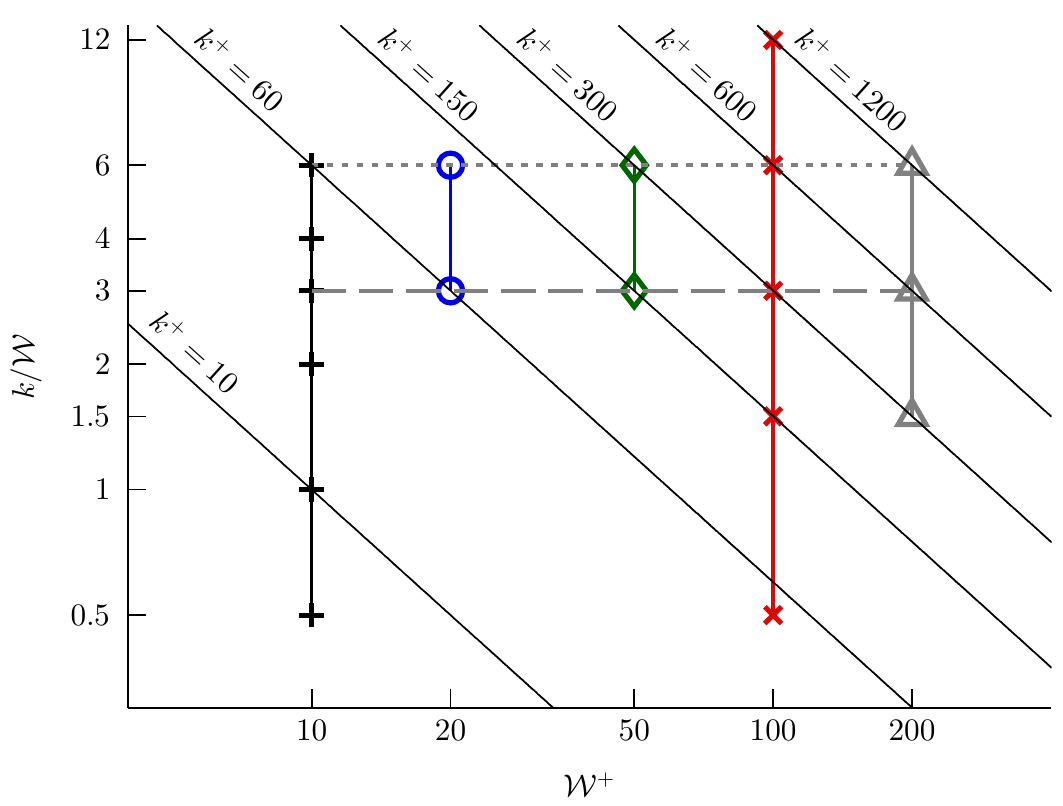}
	\put(-6.2,5.2){\vector(-1,2){1.58}}
	\put(-10.2,8.4){\includegraphics[trim=0 85 0 0,clip=true]{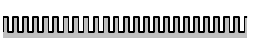}}
	\put(-2.95,6.5){\vector(4,1){2}}
	\put(-1.5,5.85){\includegraphics[trim=0 10 0 0,clip=true]{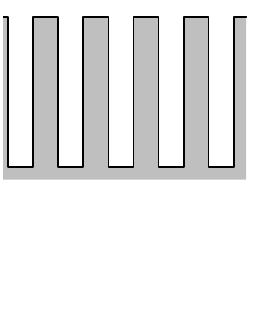}}
	\put(-2.95,1.95){\vector(1,-1){1.75}}
	\put(-1.5,-0.1){\includegraphics[trim=0 80 0 0,clip=true]{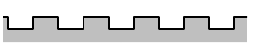}}	
	\vspace{-0.0\baselineskip}
\caption{(Colour online) Parameter space of the current study, showing the different series of spanwise bars. Fixed $\w^+$ shown by vertical solid lines, fixed $k/\w$ shown by horizontal dashed and dotted lines.
Lines of constant $k^+$ are shown by diagonal lines in this log-log plot.
A selection of roughness cross sections are indicated by the arrows.
}
\label{fig:paramSpace}
\end{figure}

Figure \ref{fig:paramSpace} shows the parameter space of different $\w^+$ values used in this study. The following figures will all use the same symbols, where data series are grouped by $\w^+$ (vertical lines in figure \ref{fig:paramSpace}). In some situations, it is useful to consider fixed $k/\w$, with this study considering two different values of $k/\w=3$ and $k/\w=6$ (horizontal dashed and dotted lines, respectively). A fixed $k/\w$ is similar to how laboratory experiments are conducted, in that a single surface (of fixed $k/\w$) is studied at different flow speeds, or $\w^+$. 
In the current simulations at fixed $Re_\tau=395$, the physical size of the bars is changed so that $\w^+$ changes.
 Also shown in figure \ref{fig:paramSpace} are lines of constant $k^+$, shown by the diagonal lines in this logarithmic plot. Studies investigating the effect of solidity, $\Lambda = k/2\w$, are typically done at constant $k^+$, whereby roughness elements with fixed height are studied with different spacings between the elements, $\w^+$ \citep[e.g.][]{Leonardi07,MacDonald16}.

\subsection{Mean flow}

\setlength{\unitlength}{1cm}
\begin{figure}
\centering

	\includegraphics{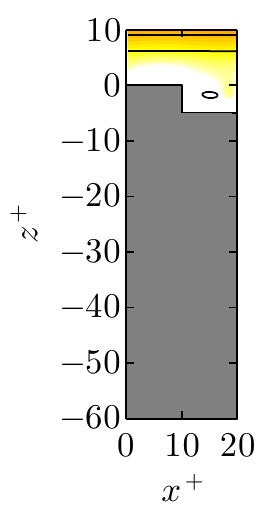}
	\includegraphics{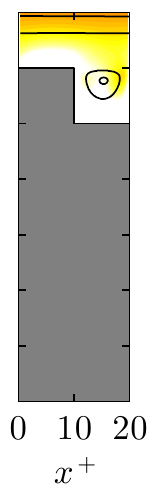}
	\includegraphics{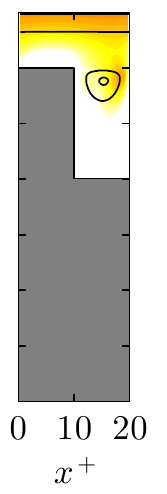}
	\includegraphics{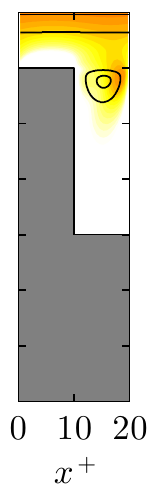}
	\includegraphics{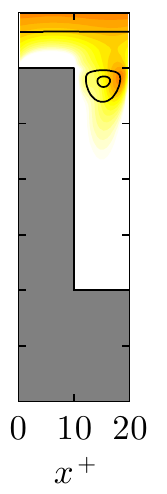}
	\includegraphics{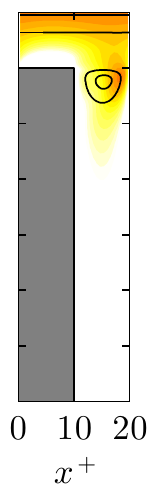}
	\put(-12.7,5){(\emph{a})}
	\put(-10.35,5){$k^+$$=5$}
	\put(-8.65,5){$k^+$$=10$}
	\put(-6.85,5){$k^+$$=20$}
	\put(-5.05,5){$k^+$$=30$}
	\put(-3.3,5){$k^+$$=40$}
	\put(-1.5,5){$k^+$$=60$}
	\vspace{0.2\baselineskip}
	\\
	\includegraphics{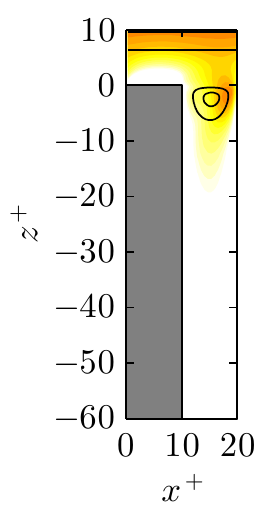}
	\hspace{-0.4cm}
	\includegraphics{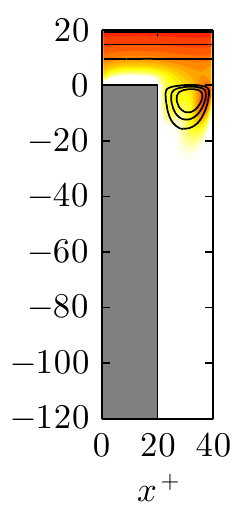}
	\hspace{-0.4cm}
	\includegraphics{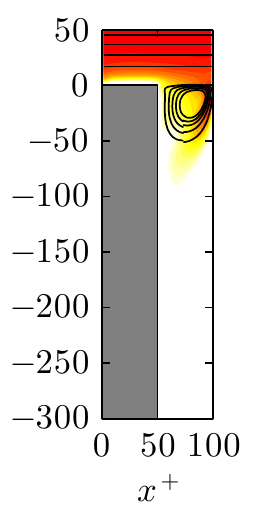}
	\hspace{-0.4cm}
	\includegraphics{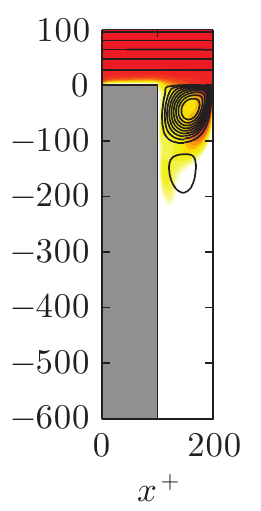}
	\hspace{-0.4cm}
	\includegraphics{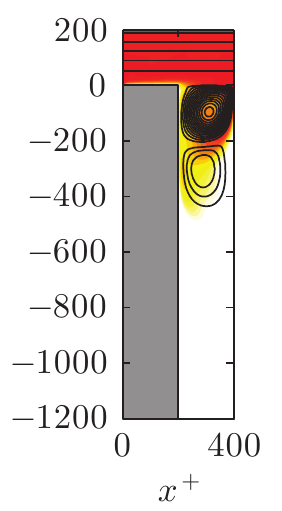}
	\put(-13.1,5){(\emph{b})}
	\put(-11.5,5){$\w^+$$=10$}
	\put(-9.1,5){$\w^+$$=20$}
	\put(-6.7,5){$\w^+$$=50$}
	\put(-4.2,5){$\w^+$$=100$}
	\put(-1.6,5){$\w^+$$=200$}
	\\
	\vspace{0.1cm}
	\includegraphics[trim=0.4cm  0.2cm  0 0]{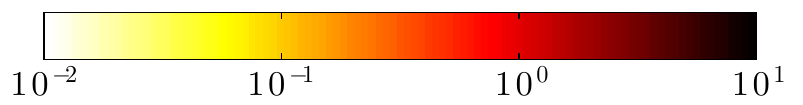}
	\put(-4.5,-0.4){${w'^+}^2$}
	\vspace{-0.0\baselineskip}
\caption{(Colour online) Streamlines of time-averaged velocity for (\emph{a}) $\w^+=10$ with $5\le k^+\le60$ and (\emph{b}) $k/\w=6$ with $10\le \w^+\le200$. Streamlines are defined such that the mass flux between each streamline is approximately constant.
Contour shows the vertical (wall-normal) turbulence intensity ${w'^+}^2$ with a logarithmic colourbar.
}
\label{fig:wContour}
\end{figure}

Streamlines of time-averaged velocity are shown in figure \ref{fig:wContour} over a colour contour of the vertical (wall-normal) turbulence intensity, ${w'^+}^2$ where the colourbar has a logarithmic scale. The intensity is  defined using the triple decomposition of $w=W+\widetilde{w}+w'$ \citep{Reynolds72,Finnigan00} where the time-independent (spatially dependent) component, $\widetilde{w}(x,z)$, of the fluctuation is subtracted off and $w'$ is the purely turbulent fluctuation.
 Firstly, we will just consider the effect of increasing $k^+$ for a fixed $\w^+=10$ (figure \ref{fig:wContour}\emph{a}).  Streamlines are defined such that the mass flow between two streamlines in the roughness cavity is approximately constant, $\int_{z_a}^{z_b} (U^+ +\widetilde{u}(x_m,z))\id z = A_s\w\approx 0.04\w$, where $x_m$ is the streamwise mid-point of the roughness cavity. 
 A clear difference for $k^+=5$ is seen, with the small roughness height restricting the recirculating `bubble'. There is a subtle difference with $k^+=10$, however once $k^+\gtrsim20$ little effect is seen with increasing $k^+$. This suggests that once $k/\w\gtrsim2$ there is very little effect in changing $k$.  At this point, there is a single recirculating bubble rotating in the clockwise direction. There is a cascade of alternately rotating recirculating regions below this primary bubble that reaches the roughness trough, however the strength of these regions are negligible  compared to the turbulent flow above, with the local velocities being less than one percent of $U_\tau$. The contour of ${w'^+}^2$ shows that the turbulent motions do not penetrate far into the roughness canopy, tending to zero within 10 to 20 viscous units.

Increasing the width of the fluid gap whilst retaining a fixed aspect ratio $k/\w$ (figure \ref{fig:wContour}\emph{b}) results in a strengthening of the fluid recirculation region, as well as a greater penetration of the turbulence into the roughness canopy in terms of $z^+$. However, it still appears that the turbulent fluctuations do not penetrate much past $z/\w\approx-2$.
 When $\w^+\ge100$, a secondary recirculating bubble emerges below the primary one based on the constant massflux definition of the streamlines defined above. This secondary recirculating bubble is nearly the same strength as the primary recirculating bubble seen when $\w^+=10$ in figure \ref{fig:wContour}(\emph{a}). This secondary bubble  is still present for  $\w^+<100$, however its strength is substantially diminished.
The streamlines in the final panel of figure \ref{fig:wContour}(\emph{b}) appear particularly dense, but are spaced approximately one grid cell apart.

\setlength{\unitlength}{1cm}
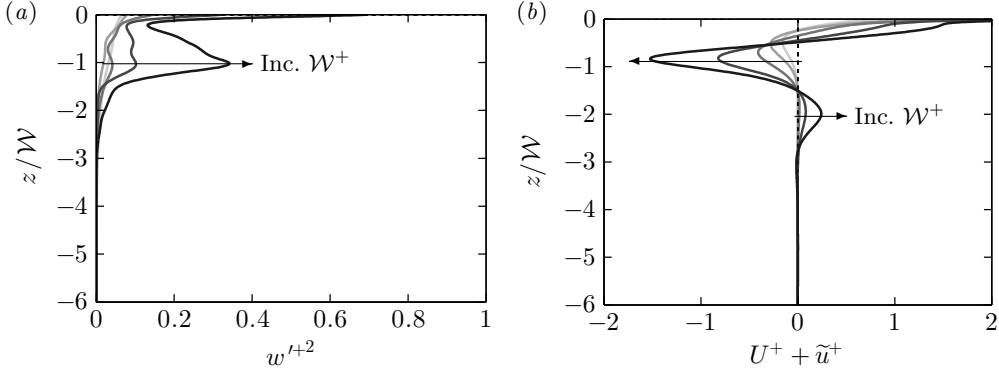
\begin{figure}
\centering
		\includegraphics[width=0.49\textwidth]{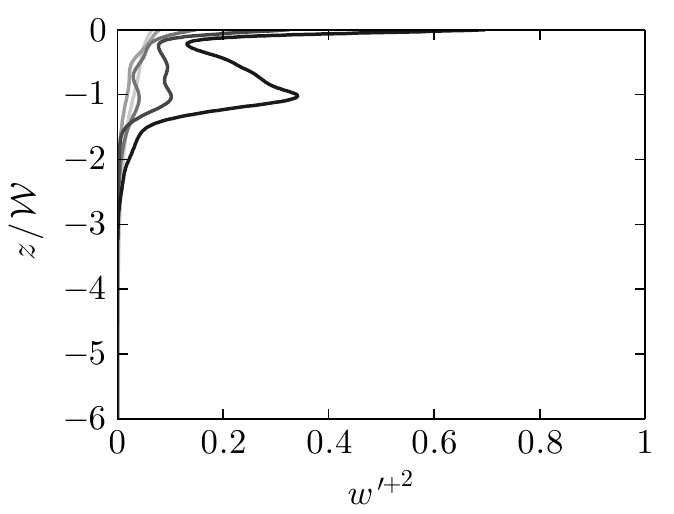}
		\includegraphics[width=0.49\textwidth]{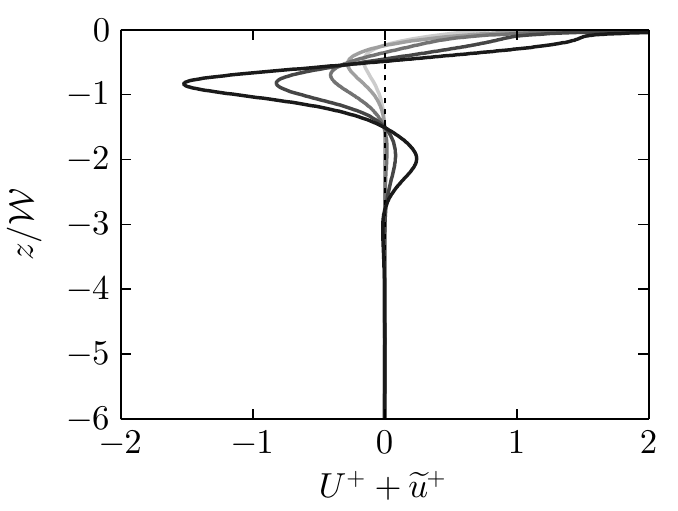}
	\put(-13.4,4.7){(\emph{a})}
	\put(-12.1,4.1){\vector(1,0){2.0}}
	\put(-10,4.0){Inc.~$\w^+$}
	\put(-6.6,4.7){(\emph{b})}
	\put(-2.8,4.13){\vector(-1,0){2.3}}
	\put(-2.9,3.4){\vector(1,0){0.7}}
	\put(-2.1,3.3){Inc.~$\w^+$}
	\vspace{-0.5\baselineskip}
\caption{Profiles of (\emph{a}) vertical turbulence intensity ${w'^+}^2$ and (\emph{b}) streamwise velocity $U^++\widetilde{u}^+$ at the streamwise mid-point of the trough, for $k/\w = 6$. Darker grey lines and arrows indicate increasing $\w^+$.
}
	\label{fig:wddProf}
\end{figure}

To better describe the flow within the cavity shown in figure \ref{fig:wContour}(\emph{b}), vertical profiles are shown in figure \ref{fig:wddProf} of the vertical turbulence intensity (\emph{a}) and streamwise velocity (\emph{b}). Note that these are not spatially averaged but are instead profiles at the streamwise mid-point of the troughs. The vertical turbulence intensity shows a clear peak forming at $z/\w=-1$ which is especially evident for the largest bar spacing of $\w^+=200$. This is associated with the bottom edge of the primary recirculating bubble observed in figure \ref{fig:wContour}(\emph{b}). Below $z/\w=-1$, the turbulence intensity quickly decreases to become negligible below $z/\w\approx-2$. The streamwise velocity, $U^++\widetilde{u}^+$ (figure \ref{fig:wddProf}\emph{b}), is negative near the bottom of the primary recirculating bubble near $z/\w=-1$, with increasing $\w^+$ resulting in stronger recirculation. The wider bar spacings of $\w^+=100$ and $\w^+=200$ also show the bottom of the secondary recirculating bubble at $z/\w\approx -2$, where a positive streamwise velocity is observed from this counter-clockwise rotating vortex. These results suggest that the strength of these recirculating bubbles continues to increase with increasing $\w^+$. While later results will suggest the turbulent flow above the roughness is close to an asymptotic fully rough state, with the mean streamwise velocity profile collapsing when plotted against $z/\w$, this does not appear to extend to the flow within the roughness cavity itself.

\setlength{\unitlength}{1cm}
\begin{figure}
\centering
		\includegraphics[width=0.49\textwidth]{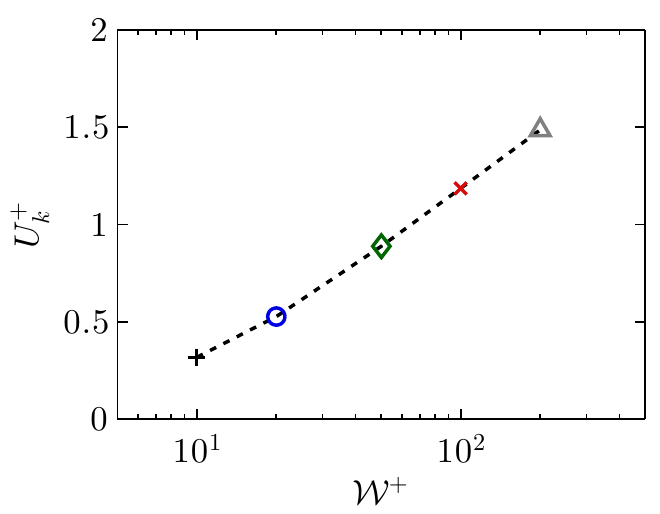}
		\includegraphics[width=0.487\textwidth]{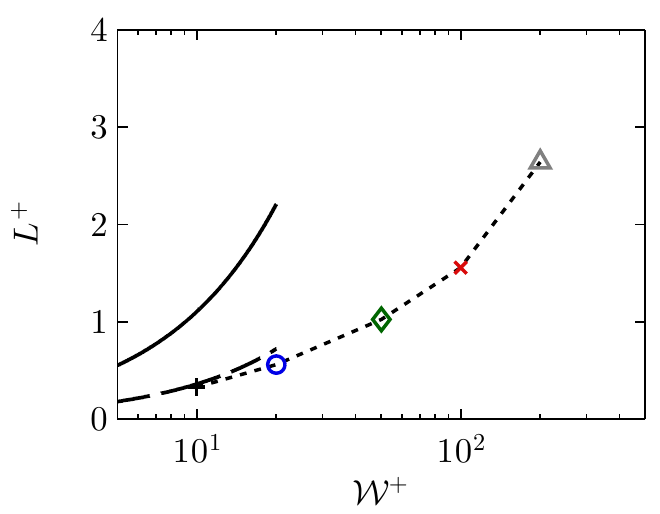}
	\put(-13.4,4.8){(\emph{a})}
	\put(-6.6,4.8){(\emph{b})}
	\put(-5.2,3.15){$L^+$$=2C_b\w^+$}
	\vspace{-0.5\baselineskip}
\caption{(Colour online) (\emph{a}) Slip velocity at the roughness crest $U_k^+$ and (\emph{b}) slip length $L^+$ for varying fluid widths $\w^+$.
Data are for fixed $k/\w=6$.
Solid line in (\emph{b}) indicates the Stokes flow limit $L^+=2C_b \w^+$ for alternating slip/no-slip surfaces \citep{Lauga03}, with $C_b=0.055$ for $\w/\lambda=0.5$.
Dashed line indicates a fit of $C_b\approx0.018$ for the present data.
}
	\label{fig:slipLength}
\end{figure}

The  streamlines of figure \ref{fig:wContour} suggest that, in the mean, the flow above the roughness sees an alternating slip/no-slip interface at the roughness crest. This scenario is often used in the superhydrophobic surface literature to describe the flow over textured surfaces similar to the present ones, in which bubbles of gas are trapped inside the roughness canopy with a liquid flowing over the top. The solid posts present a no-slip interface to the liquid, while the gas presents a zero shear interface which can promote drag reduction. This is similar to the present flow, with the notable exception of zero permeability in the region between the bars. A key parameter in the study of superhydrophobic surfaces is the slip length, $L^+$, which relates the slip velocity at the roughness crest, $U_k^+$ (averaged over both slip and no-slip surfaces) to the average velocity gradient at the crest, $U^+_k=L^+ {\dd U_k^+}/{\dd z^+}$. 

The slip velocity is shown in figure \ref{fig:slipLength}(\emph{a}) for $k/\w=6$ and can be seen to increase with $\w^+$. 
 The slip length in figure \ref{fig:slipLength}(\emph{b}) also increases with $\w^+$, in agreement with previous studies of turbulent flow over superhydrophobic surfaces comprising square posts \citep{Seo16}. Here, \cite{Seo16} suggested that in the limit of $\w^+\rightarrow0$ the slip length approaches the Stokes flow solution for flow over a slip/no-slip wall. \cite{Lauga03} determined that the asymptotic limit for Stokes flow over spanwise aligned textures with alternating slip/no-slip boundary conditions was $L^+=2C_b(\sigma)\w^+$, where $\sigma=\w/\lambda$ is the fraction of fluid area at the crest and the coefficient $C_b=-\log(\cos(\sigma\pi/2))/2\pi$. This is displayed in figure \ref{fig:slipLength}(\emph{b}) with $C_b(\sigma=0.5)=0.055$, where the present turbulent flow data show a smaller slip length to that predicted by this Stokes flow solution with the constant estimated as $C_b\approx0.018$. This is expected as the gaps between the spanwise bars are permeable and are not a pure slip boundary, as assumed in the Stokes flow case. However, these similarities to superhydrophobic surfaces (which neglect any flow within the roughness canopy) may suggest there is little interaction between the roughness canopy flow and the flow above the roughness.
 
\setlength{\unitlength}{1cm}
\begin{figure}
\centering
	\includegraphics[width=0.49\textwidth]{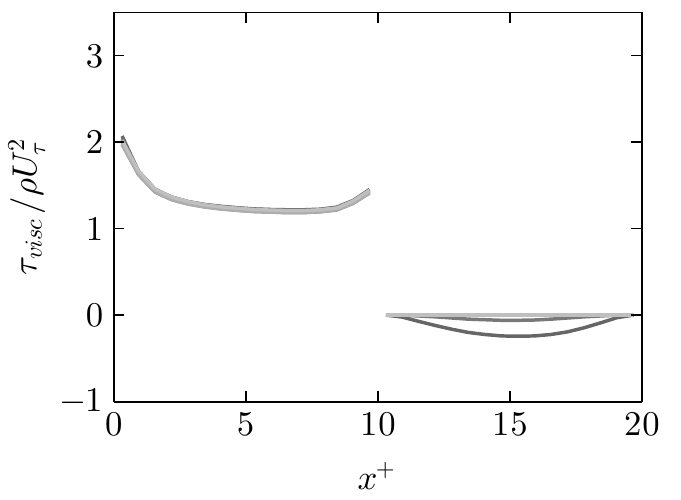}
	\includegraphics[width=0.49\textwidth]{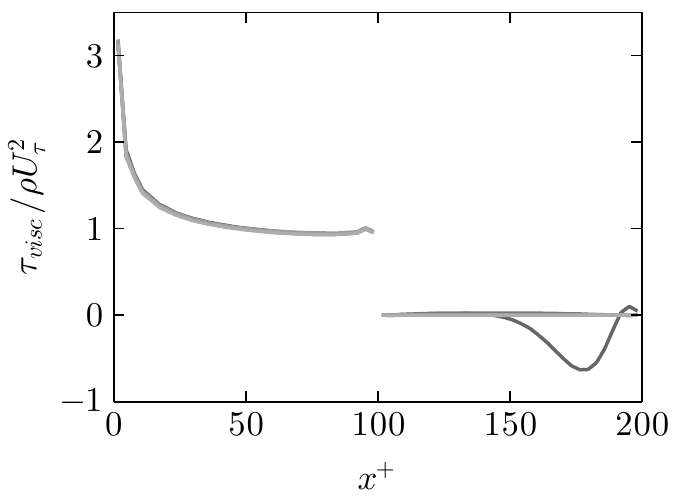}
	\put(-13.6,4.75){(\emph{a})}
	\put(-6.8,4.75){(\emph{b})}
	\put(-10.4,1.2){$k^+$$=5$}
	\put(-9.45,1.3){\line(2,1){0.63}}
	\put(-8.3,1.1){$k^+$$=10$}
	\put(-8.35,1.2){\line(-1,4){0.14}}
	\put(-10.5,1.9){$k^+$$\ge20$}	
	\put(-9.4,1.97){\line(3,-1){0.56}}
	\put(-11.5,3.0){Crest}
	\put(-8.7,1.95){Trough}
	\put(-4.6,2.8){Crest}
	\put(-1.9,1.95){Trough}
	\put(-2.9,1.2){$k^+$$=50$}
	\put(-1.8,1.3){\line(4,1){0.55}}
	\put(-3.5,1.9){$k^+$$\ge150$}	
	\put(-2.23,1.97){\line(3,-1){0.53}}
	\vspace{-0.5\baselineskip}
	\\
	\includegraphics[width=0.49\textwidth]{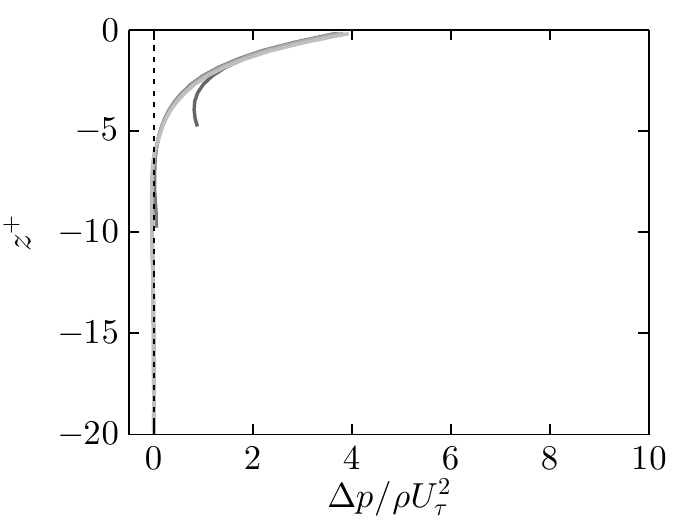}
	\includegraphics[width=0.49\textwidth]{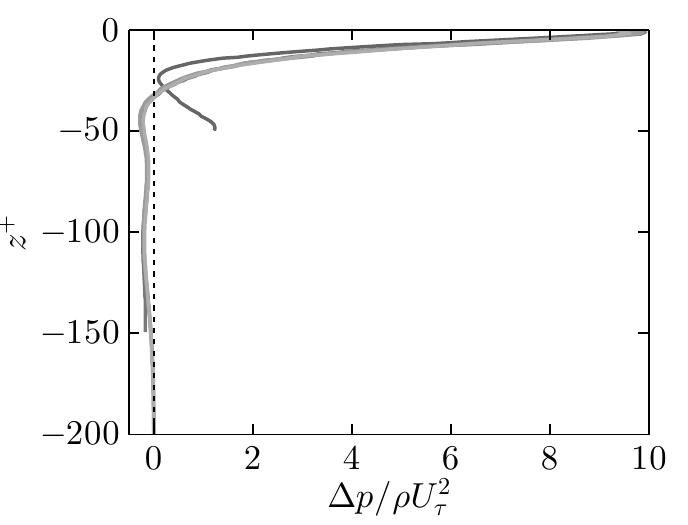}
	\put(-13.6,4.75){(\emph{c})}
	\put(-6.8,4.75){(\emph{d})}
	\put(-10.2,4.35){Crest}
	\put(-11.35,3.75){$k^+=5$}
	\put(-11.75,2.77){$k^+=10$}
	\put(-11.75,1.0){$k^+\ge 20$}
	\put(-1.4,4.35){Crest}
	\put(-4.5,3.72){$k^+=50$}
	\put(-5.05,1.80){$k^+=150$}
	\put(-5,1.0){$k^+\ge 300$}
	\vspace{-0.5\baselineskip}
\caption{
(\emph{a,b}) Wall shear stress against streamwise position and (\emph{c,d}) total pressure difference across a single roughness element against vertical position ($z^+=0$ is located at the roughness crest). Distance between bars of (\emph{a},\emph{c}) $\w^+\approx 10$ and (\emph{b},\emph{d}) $\w^+\approx 100$. Lighter grey denotes larger $k^+$ (table \ref{tab:sims}).
}
	\label{fig:stressDist}
\end{figure}
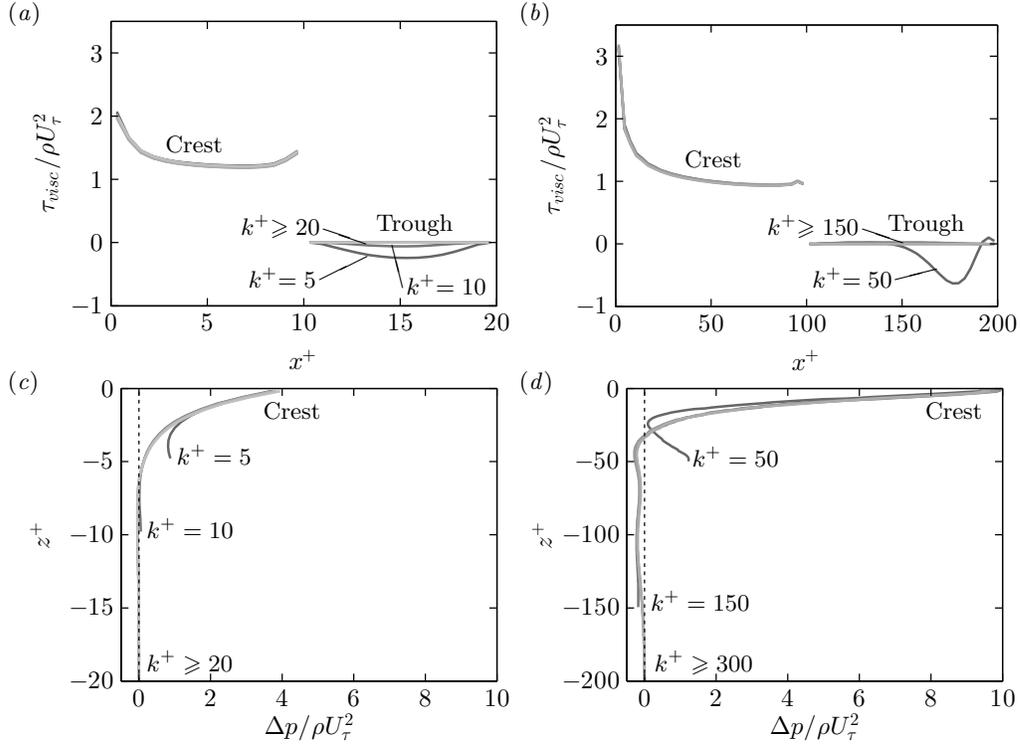

Figure \ref{fig:stressDist} shows the time-averaged viscous stress and pressure distribution acting on a single roughness element for cases with $\w^+\approx10$ and $\w^+\approx100$. The viscous stress, $\tau_{visc}^+(x)$, only acts on the horizontal roughness crests and troughs and can be related to the viscous drag force via $F_\nu=\int_0^\lambda \tau_{visc} \id x\cdot L_y L_x/\lambda$. The viscous stress distribution along the roughness crest does not vary with $k$ for either $\w^+\approx10$ (\emph{a}) or $\w^+\approx100$ (\emph{b}). There is some variation in the viscous stress at the trough, whereby the shortest roughness heights have a negative stress due to recirculation within the roughness cavity. However, once $k/\w\gtrsim1$ the trough is sufficiently removed from this recirculation element such that the viscous stress is approximately zero.

The total pressure distribution, $\Delta p^+(z)$ is shown for $\w^+\approx10$ and $\w^+\approx100$ in figure \ref{fig:stressDist}(\emph{c}) and (\emph{d}). As discussed in \S\ref{sect:vol}, the pressure across a single element, $\Delta p^+$, is the combination of the driving (mean) pressure $\Delta P^+$ and the periodic component $\Delta \pp^+$. The total  pressure drag force in the channel is then $F_{p}=\int_{-kt}^0\Delta p\id z\cdot L_y L_x/\lambda$.  This pressure difference can be seen to tend to zero when $k/\w\gtrsim1$ for $\w^+\approx 10$ (\emph{c}). However for the  roughness with $\w^+\approx100$ (\emph{d}) there is a slight negative pressure difference below $z^+\approx -30$, before tending to zero at $z^+\approx-150$. This is likely due to the greater strength of the secondary vortex seen over $-200\lesssim z^+\lesssim -100$ for $\w^+=100$ (figure \ref{fig:wContour}\emph{b})

\setlength{\unitlength}{1cm}
\begin{figure}
\centering
	\includegraphics[width=0.65\textwidth]{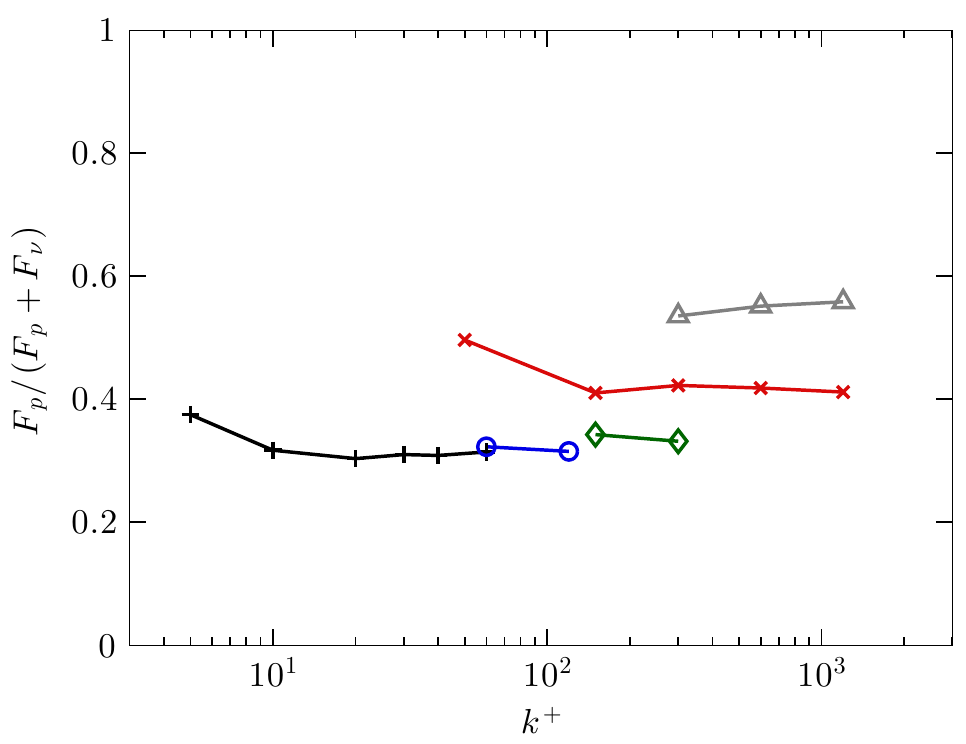}
	\vspace{-0.2\baselineskip}
\caption{(Colour online) Ratio of total pressure drag force, $F_{p}$, to the total drag force.
Symbols:
\protect\raisebox{0.4ex}{\protect\scalebox{1.0}{$\color{black}\boldsymbol{\pmb{+}}$}}, $\w^+=10$;
\protect\raisebox{0.15ex}{\protect\scalebox{1.3}{$\color{myblue}\boldsymbol{\pmb{\circ}}$}}, $\w^+=20$;
\protect\raisebox{0.3ex}{\protect\scalebox{1.0}{$\color{mygreen}\boldsymbol{\pmb{\diamondsuit}}$}}, $\w^+=50$;
\protect\raisebox{0.3ex}{\protect\scalebox{1.0}{$\color{myred}\boldsymbol{\pmb{\times}}$}}, $\w^+=100$;
\protect\raisebox{0.1ex}{\protect\scalebox{1.0}{$\color{LGrey}\boldsymbol{\pmb{\bigtriangleup}}$}}, $\w^+=200$.
}
	\label{fig:FPratio}
\end{figure}

Figure \ref{fig:FPratio} shows the ratio of the total pressure drag force $F_{p}$ to the overall drag force, $D_1 = F_{p}+F_\nu$. For the three narrowest gaps of $\w^+\lesssim50$, the pressure drag is approximately 32\% of the total drag and is almost independent of $k$. However, the larger widths have larger pressure drag contributions, of 42\% of the total drag for $\w^+=100$ and 57\% for $\w^+=200$. These magnitudes are similar to those observed in three-dimensional sinusoidal roughness which exhibit $k$-type behaviour \citep{Chan15,MacDonald16}.  This increase in pressure drag contribution with increasing $\w^+$ is due to the increased turbulence penetration and strengthening of the recirculating bubbles that was observed in figure \ref{fig:wContour}.
In the seminal $d$-type work of \cite{Perry69}, the authors assumed that the viscous drag was negligible so that the pressure drag, which could be experimentally measured using pressure tappings, was the only contribution to $U_\tau$. While this assumption was for bars with $\w^+\gtrsim250$, the present data for $\w^+\approx 200$ shows that the viscous drag remains at approximately 40\% of the total drag, suggesting that the assumption of negligible viscous drag may not be ideal.

\setlength{\unitlength}{1cm}
\begin{figure}
\centering
	\includegraphics[width=0.49\textwidth]{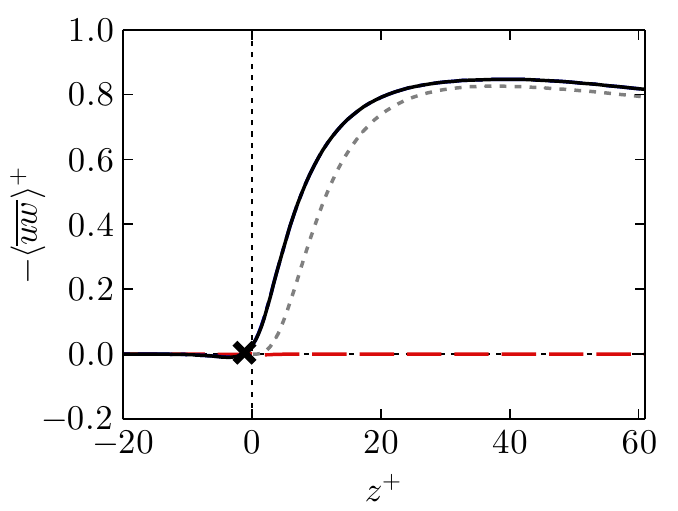}
	\includegraphics[width=0.48\textwidth]{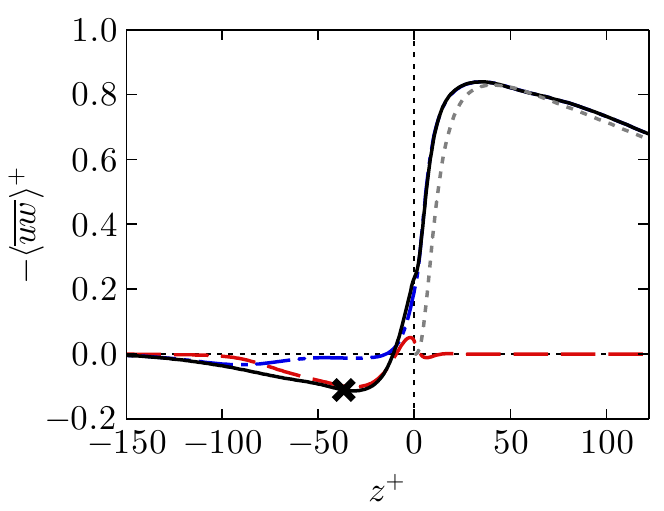}
	\put(-13.2,4.7){(\emph{a})}
	\put(-6.4,4.7){(\emph{b})}
	\vspace{-0.5\baselineskip}
\caption{(Colour online) Reynolds stress decomposed into dispersive and turbulent components for $k/\w=6$, with (\emph{a}) $\w^+=10$ and (\emph{b}) $\w^+=100$. Line styles:
\solidLine{black}, total Reynolds stress (dispersive $+$ turbulent);
\dashLine{myred}, dispersive stress $\langle\widetilde{u}\widetilde{w}\rangle^+$;
\dashDotLine{myblue}, turbulent stress $\langle \overline{u'w'}\rangle^+$;
\dotLine{LGrey}, smooth-wall total Reynolds stress.
Cross denotes virtual origin derived from the force-centroid method \citep{Jackson81}.
Origin in $z^+$ is the roughness crest.
}
	\label{fig:dispStress}
\end{figure}

The Reynolds stress for rough-wall flows can be decomposed into two components based on the triple decomposition defined above \citep{Reynolds72,Finnigan00}. These are the time-independent (spatially dependent) component, known as the dispersive stress $-\langle\widetilde{u}\widetilde{w}\rangle^+$ and the purely turbulent (time dependent) stress $-\langle \overline{u' w'}\rangle$. Here, the overbar denotes temporal averaging, with $\overline{\widetilde{u}}=\widetilde{u}$, while the angle brackets denote the superficial spatial average over a wall-parallel volume which is thin in the vertical direction. This superficial average includes both fluid and solid regions, where the solid region has zero velocity everywhere. This can be related to the intrinsic spatial average (averaging over just the fluid regions) by multiplying the superficially averaged quantity by the ratio of total volume to fluid-only volume, in this case $\lambda/\w=2$. The distinction between superficial and intrinsic averaging becomes irrelevant above the roughness crests ($z>0$) as there is only fluid region in the wall-parallel plane.
The cross terms involving $u_i'$ and $\widetilde{u}_i$ will be zero, because $\overline{u_i'}=0$ and $\langle \widetilde{u}_i\rangle=0$.
Figure \ref{fig:dispStress} shows these two components, along with the total Reynolds stress (sum of dispersive and turbulent stresses). For the narrow bars with $\w^+=10$ (figure \ref{fig:dispStress}\emph{a}), the dispersive stress is negligible between the bars, and zero above the roughness crest. The total Reynolds stress is therefore essentially the same as the turbulent stress. However, compared to the smooth-wall Reynolds stress (dotted line), the rough-wall total Reynolds stress appears to have a different virtual origin that is within the roughness canopy.

Figure \ref{fig:dispStress}(\emph{b}) shows that the dispersive stress (dashed line) for the wider bars ($\w^+=100$) is dominant and counter-gradient within the roughness canopy, while the turbulent stress is near zero in this region. This strong dispersive stress (relative to the total)  would be due to the recirculating bubbles within the canopy, which due to the time-independent nature of the dispersive stress suggest that these bubbles do not vary much with time. Moreover, the near-zero turbulent Reynolds stress suggests that even though vertical turbulent fluctuations penetrate into some of the canopy (contours in figure \ref{fig:wContour}), this does not result in significant momentum flux through the roughness canopy. Above the roughness crest, the dispersive stress quickly tends to zero as the flow retains little spatially dependent information of the bars. Note that this does not imply the flow has lost all information regarding the bars; rather it could have temporally dependent eddies that would not appear as a time-independent dispersive stress. As with the narrower bars (figure \ref{fig:dispStress}\emph{a}), the total Reynolds stress appears to have a virtual origin that is within the roughness canopy.

The dispersive stress has often been assumed to be negligible relative to the total turbulent Reynolds stress.
This is based on laboratory experiments such as those of \cite{Poggi04} and \cite{Poggi08}, who determined that the dispersive stress became negligible for three-dimensional vertical cylinders when $\Lambda>0.1$. Similarly, \cite{Bohm13} showed that the dispersive stress is smaller than the turbulent stress for a three-dimensional array of bulbs with $\Lambda=0.38$.  However, a recent analysis by \cite{Castro17exp} on the numerical studies of \cite{Leonardi10} and \cite{Castro17} showed that the dispersive stress is significant for cuboid roughnesses, supporting the present results.

\subsection{Virtual origin}
\label{ssect:origin}

\setlength{\unitlength}{1cm}
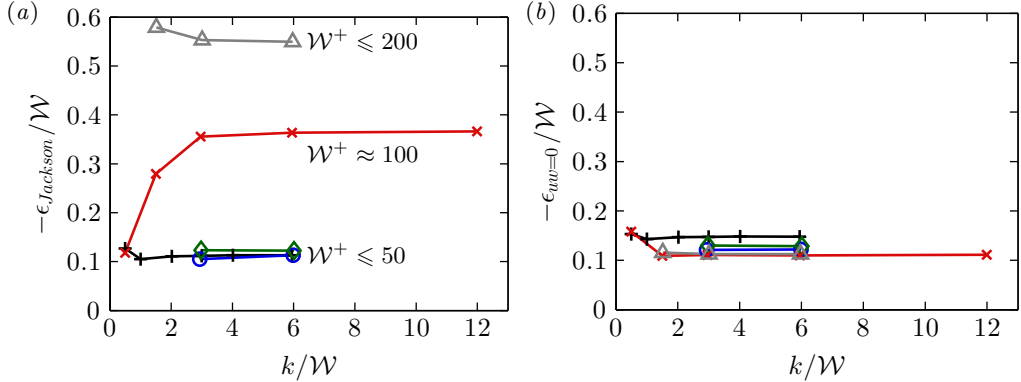
\begin{figure}
\centering
	\includegraphics[width=0.49\textwidth]{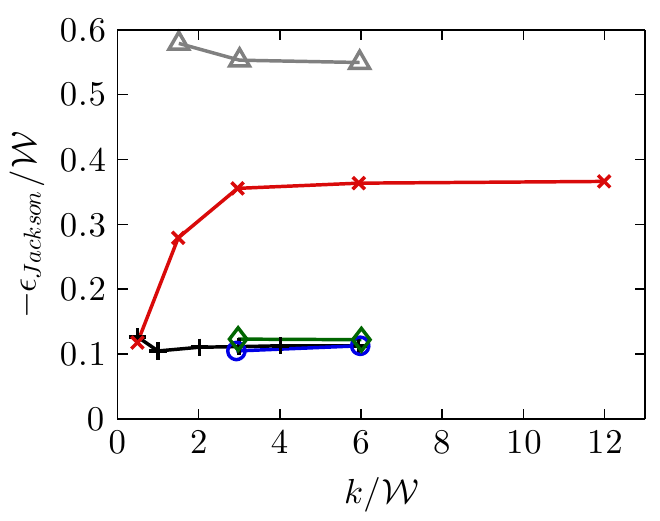}
	\includegraphics[width=0.495\textwidth]{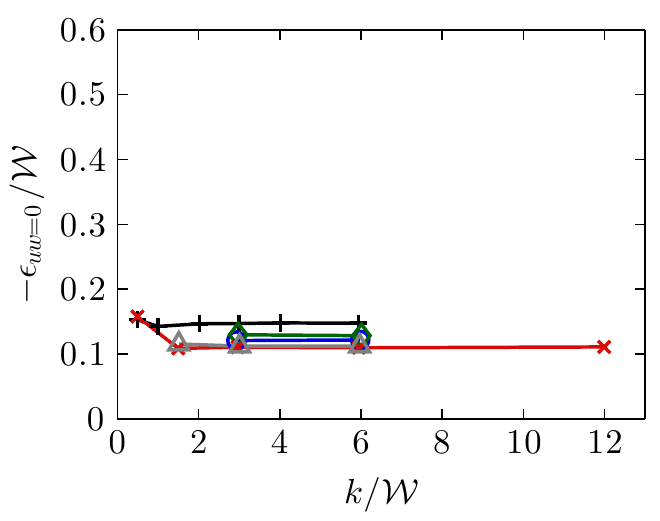}
	\put(-13.6,4.9){(\emph{a})}
	\put(-9.6,1.65){$\w^+\le50$}
	\put(-9.6,3.0){$\w^+\approx100$}
	\put(-9.6,4.5){$\w^+\le200$}	
	\put(-6.7,4.9){(\emph{b})}
	\vspace{-0.2\baselineskip}
	\vspace{-0.2\baselineskip}
\caption{(Colour online) Virtual origin, calculated using 
(\emph{a}) the force-centroid method \citep{Jackson81}
and 
(\emph{b}) the location of zero Reynolds stress.
Symbols are the same as figure \ref{fig:paramSpace}.
}
	\label{fig:origin}
\end{figure}

A common issue in roughness studies is determining the location of the origin in the vertical coordinate $z$, often termed the zero-plane displacement height or the virtual origin, $\epsilon$. 
The zero-plane displacement height, often denoted by $d$, is a dynamic parameter associated with the origin of the logarithmic region of the flow, given by $U^+=\kappa^{-1}\log(z-d)+B_s-\Delta U^+$. This is therefore found by fitting the log-law to the mean velocity profile, although this can be problematic as the roughness function $\Delta U^+$ is also unknown. Furthermore, for the present simulations, $Re_\tau$ is relatively small and the use of the minimal channel means that only the beginning of the logarithmic region is captured. This makes it difficult to fit a log law to the velocity profile so that the zero-plane displacement height cannot be easily measured in this way.
The term virtual origin is often used interchangeably with the zero-plane displacement height, but typically refers to methods that estimate the origin based on flow parameters other than the logarithmic velocity profile.

In the limit of tightly packed bars ($\w^+\ll 1$), the roughness is submerged in a viscous-dominated flow while the turbulent flow above appears as a time-dependent shear. The small-scale roughness alters the near-wall flow by promoting a slip velocity at the crests, but otherwise the outer shear is unaffected apart from the slip velocity. This idea was successfully used by \cite{Luchini91} for riblets, where the difference in streamwise and spanwise shear flows helps explain the drag-reducing properties of these small-scale protrusions. However, using the slip velocity ($U_k^+$, shown in figure \ref{fig:slipLength}\emph{a}) to define the origin breaks down for larger $\w^+$ as inertial effects begin to dominate and the assumption of viscous-dominated flow can no longer apply.

 A physically appealing method that is often employed is to define $\epsilon$ as the centroid of the moment of the drag forces acting on the rough wall \citep{Jackson81}.
Figure \ref{fig:origin}(\emph{a}) shows the virtual origin, calculated using Jackson's method. The origin for the three smallest $\w^+$ values is almost invariant with $k$, with $\epsilon\approx0.11\w$. Note that this means $\epsilon^+$ increases with $\w^+$ however the collapse with the scaling of $\epsilon/\w$ suggests $\w$ is a crucial parameter for the present roughness.
For the larger values of $\w^+\ge 100$, $\epsilon$ is substantially larger than the narrower cases, however still appears to asymptote to constant values for $k/\w\ge 3$. These constants are approximately $\epsilon\approx0.36\w$ for $\w^+=100$, and $\epsilon\approx0.55\w$ for $\w^+=200$.  This increase with $\w^+$ is most likely due to the secondary recirculating bubble observed in figure \ref{fig:wContour}(\emph{b}), which exists over $-200\lesssim z^+\lesssim -100$. This recirculating bubble causes a non-negligible pressure difference to act across the roughness elements (figure \ref{fig:stressDist}\emph{d}) which in turn causes $\epsilon^+$ to increase. Given that this secondary recirculating bubble extends down to $z^+\approx200$, then it follows we require at least $k^+>200\Rightarrow k /\w>2$ to reach this asymptote. These results are in qualitative agreement with the high aspect ratio model of \cite{Sadique17}, based on flow over three-dimensional rectangular prisms, which predicts $\epsilon/\w$ to asymptote to a constant for $k/\w\gtrsim 5$. It is likely that the weaker sheltering behaviour of three-dimensional roughness \citep{Yang16} is what requires a larger $k/\w$ ratio to reach the asymptotic limit, compared to $k/\w\gtrsim2$ observed for the present two-dimensional roughness.

This virtual origin using Jackson's method was shown in figure \ref{fig:dispStress} as a cross symbol on the total Reynolds shear stress. This is seen to be located close to the position of the strongest dispersive stress, with the other $\w^+$ cases (not shown) supporting this observation. This supports the view that the drag centroid virtual origin is related to the recirculating bubbles. However, the turbulent flow above the roughness is unlikely to depend on these temporally independent recirculating bubbles. Instead, perhaps a better location to use as the origin is where the Reynolds shear stress reaches zero \citep{Nepf07,Nepf12}. This would lead to a better collapse of the smooth- and rough-wall total Reynolds shear stress in the near-wall region as they will both reach zero at the same location.
In the present roughness, both the turbulent and dispersive Reynolds shear stress reach zero near the same location (see figure \ref{fig:dispStress}(\emph{b}) for an example with $\w^+=100$). Figure \ref{fig:origin}(\emph{b}) shows the location of the zero crossing of the turbulent Reynolds stress, $\langle\overline{u'w'}\rangle^+$. This shows a much better collapse for the larger $\w^+$ values, with $\epsilon\approx0.11\w$ for most of the cases with $\w^+\ge 20$. It is only the narrowest case with $\w^+=10$ that the Reynolds stress reaches zero slightly deeper into the troughs with $\epsilon\approx0.15\w$, although given $\w^+$ is so small then this is a minor difference in terms of viscous units. This Reynolds stress zero crossing will be used to define the origin when computing the roughness function in \S\ref{ssect:du}, although using a different origin only changes the numerical values of $\Delta U^+$ and not the conclusions of this paper.

\subsection{Mean velocity and roughness function}
\label{ssect:du}
\setlength{\unitlength}{1cm}
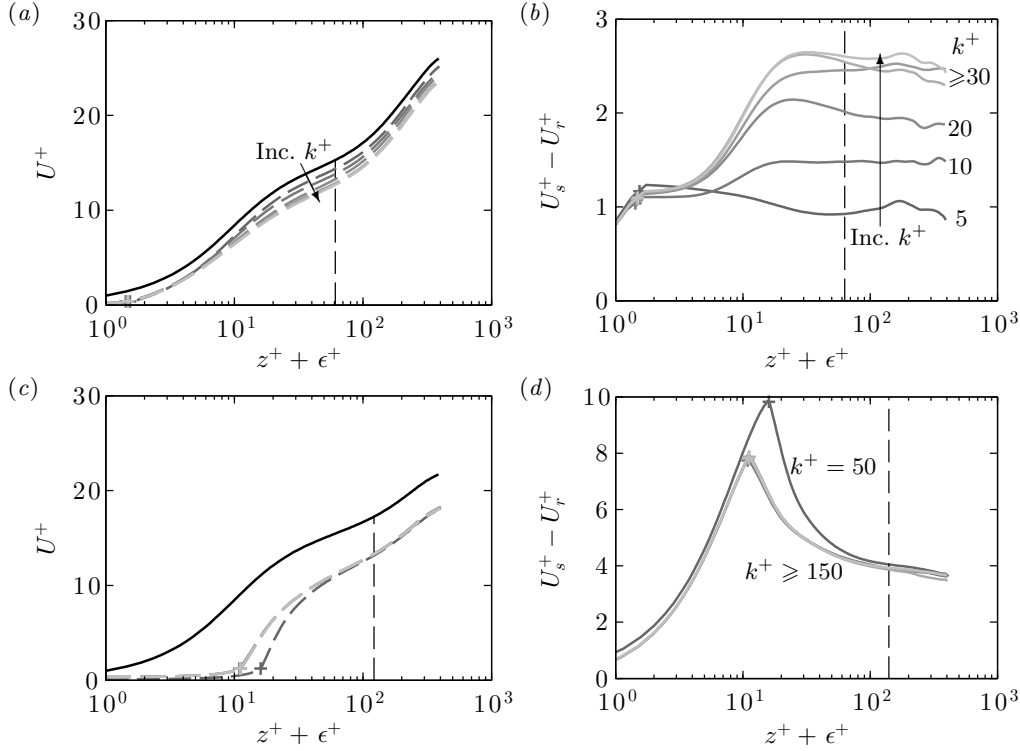
\begin{figure}
\centering
		\includegraphics[width=0.49\textwidth]{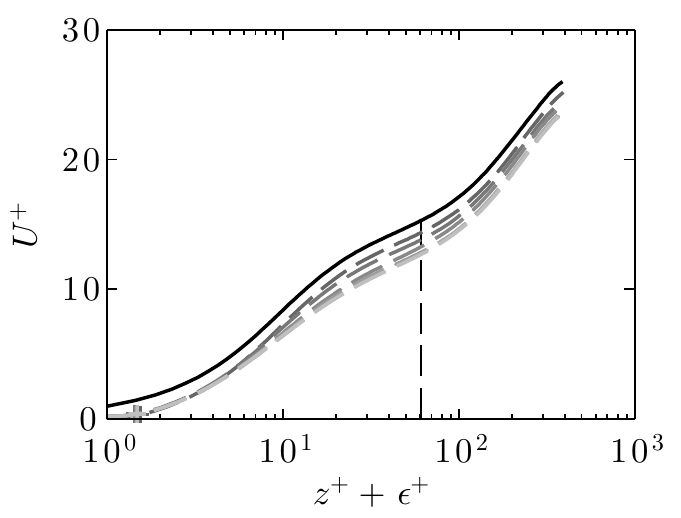}
		\includegraphics[width=0.49\textwidth]{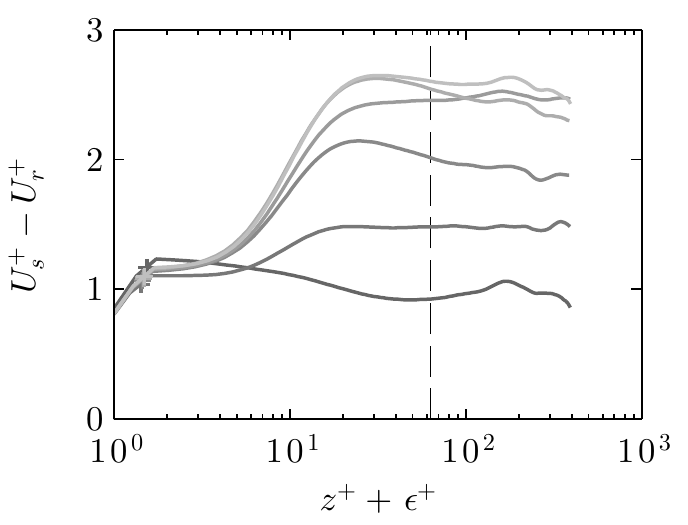}
	\put(-13.6,4.75){(\emph{a})}
	\put(-9.7,2.83){\vector(1,-2){0.25}}
	\put(-10.3,2.9){Inc. $k^+$}
	\put(-6.8,4.75){(\emph{b})}
	\put(-2,2){\vector(0,1){2.3}}
	\put(-2.4,1.75){Inc. $k^+$}
	\put(-1.0,2.0){$5$}
	\put(-1.1,2.7){$10$}
	\put(-1.1,3.2){$20$}
	\put(-1.1,3.9){$\ge$$30$}
	\put(-1.05,4.3){$k^+$}
	\vspace{-0.2\baselineskip}
	\\
		\includegraphics[width=0.49\textwidth]{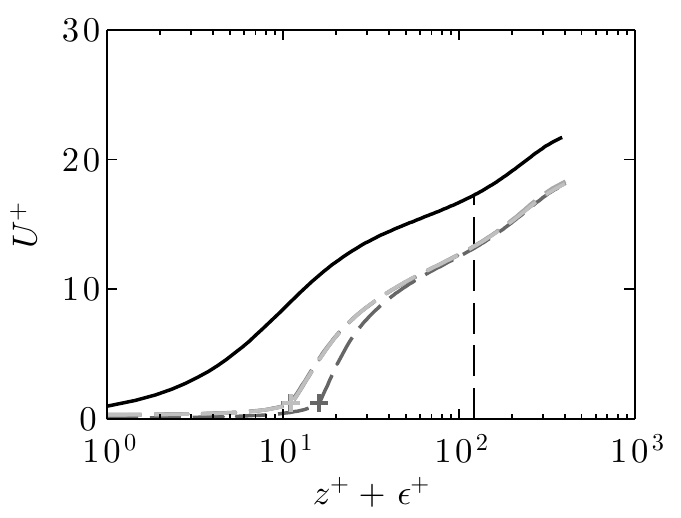}
		\includegraphics[width=0.49\textwidth]{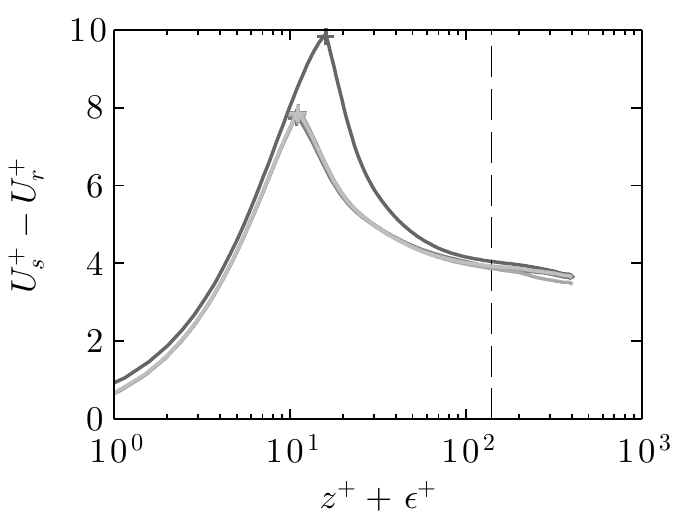}
	\put(-13.6,4.75){(\emph{c})}
	\put(-6.8,4.75){(\emph{d})}
	\put(-3.2,3.7){$k^+=50$}
	\put(-3.8,2.3){$k^+\ge 150$}
	\vspace{-0.5\baselineskip}
\caption{
(\emph{a},\emph{c})
Mean velocity profile  and
(\emph{b},\emph{d}) difference in smooth- and rough-wall velocities for fluid width (\emph{a},\emph{b}) $\w^+=10$ and (\emph{c},\emph{d}) $\w^+=100$.
Line styles: black, smooth wall; grey, rough wall.
Lighter grey denotes larger $k^+$ (table \ref{tab:sims}).
Arrow shows increasing $k^+$. 
Roughness crest denoted by $+$ symbol, origin $\epsilon^+$ defined using the Reynolds stress zero crossing (figure \ref{fig:origin}\emph{b}).
Vertical dashed line shows the minimal-span critical height $z_c^+\approx0.4L_y^+$.
}
	\label{fig:Velwp10}
\end{figure}

The mean velocity profile is shown in figure \ref{fig:Velwp10} for the cases with $\w^+=10$  and $\w^+=100$. The velocity difference between smooth- and rough-wall flows for $\w^+=10$ (figure \ref{fig:Velwp10}\emph{b}) has a dependence on $k$ for $k^+\lesssim 20$, wherein increasing $k^+$ increases the velocity difference but when $k$ is sufficiently large ($k^+\gtrsim 30\Rightarrow k/\w\gtrsim3$) little difference is observed. Similarly, for $\w^+=100$ a collapse is observed for $k/\w\gtrsim3$ (figure \ref{fig:Velwp10}\emph{c},\emph{d}), where the cases with the largest peak-to-trough heights ($k^+=300$, $600$ and $1200$) are almost indistinguishable from one another. Here, the case with the smallest roughness height of $k^+=50$ has a velocity difference slightly greater then the larger $k^+$ values, however this is likely due to differences in the virtual origin (figure \ref{fig:origin}\emph{b}).

\setlength{\unitlength}{1cm}
\begin{figure}
\centering

	\includegraphics[width=0.65\textwidth]{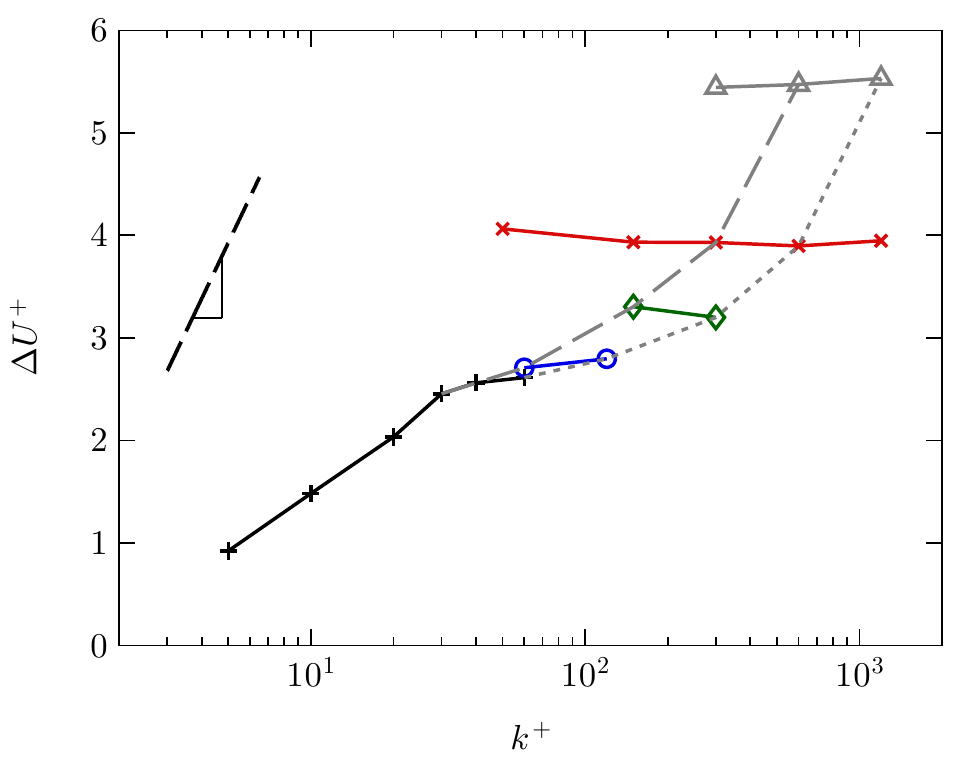}
	\put(-8.6,6.6){(\emph{a})}
	\put(-6.65,4.15){$\frac{1}{\kappa}\log(k^+)$}
	\put(-6.2,2.0){\scriptsize{\color{black}$\w^+$=$10$}}
	\put(-3.0,3.5){\scriptsize{\color{myblue}$\w^+$=$20$}}
	\put(-2.02,3.95){\scriptsize{\color{mygreen}$\w^+$=$50$}}
	\put(-4.1,4.95){\scriptsize{\color{myred}$\w^+$=$100$}}
	\put(-2.2,6.35){\scriptsize{\color{LGrey}$\w^+$=$200$}}
	\put(-2.35,4.95){\scriptsize \rotatebox{65}{$k/\w=3$}}
	\put(-1.6,4.9){\scriptsize \rotatebox{65}{$k/\w=6$}}
	\\
	\vspace{-0.3\baselineskip}
	\includegraphics[width=0.65\textwidth]{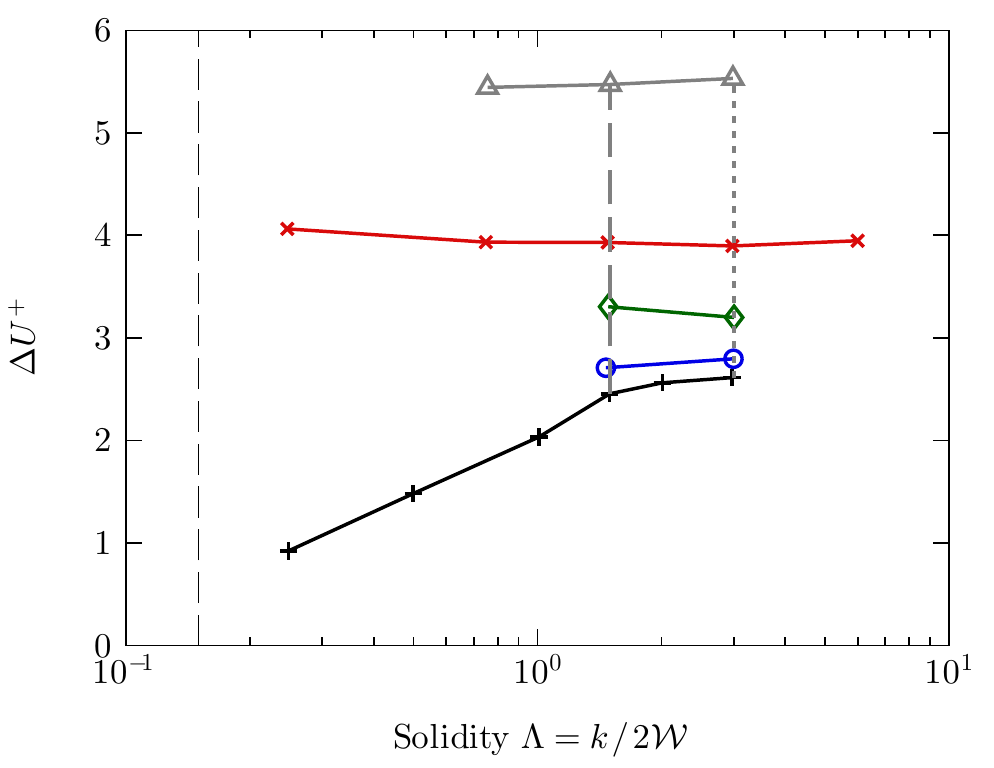}	
	\put(-8.6,6.45){(\emph{b})}
	\vspace{-0.5\baselineskip}
\caption{(Colour online) Roughness function $\Delta U^+$ against (\emph{a}) roughness trough-to-peak height $k^+$ and (\emph{b}) solidity $\Lambda=k/2\w$.
Symbols are the same as figure \ref{fig:paramSpace}.
Vertical black dashed line at $\Lambda=0.15$ in (\emph{b}) denotes the dense regime.
Vertical grey dashed and dotted lines show fixed $\Lambda=1.5$ and $\Lambda=3.0$ and are plotted in figure \ref{fig:DUw} as a function of $\w^+$.
}
\label{fig:DU}
\end{figure}
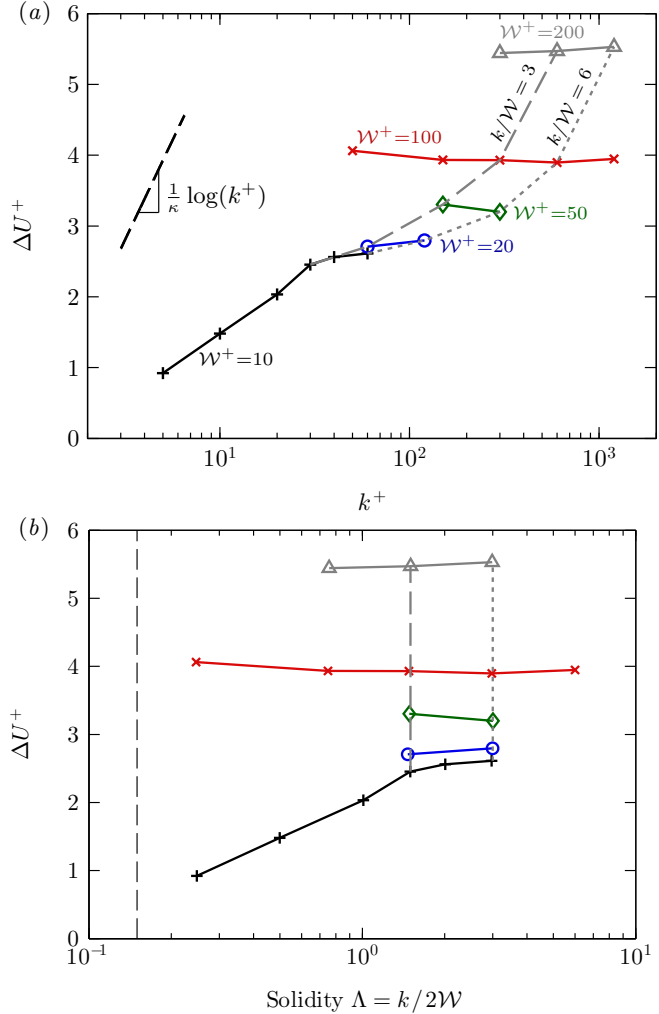

The roughness function $\Delta U^+$ is computed by taking the value of $U_{s}^+-U_{r}^+$ at the vertical critical height $z_c^+\approx0.4 L_y^+$, shown by the vertical dashed line in figure \ref{fig:Velwp10}(\emph{b},\emph{d}). Figure \ref{fig:DU}(\emph{a}) shows the roughness function against the peak-to-trough height, $k^+$. It is clear that increasing $k^+$ for fixed values of $\w^+$ does not tend towards the fully rough asymptote of $k$-type roughness, that is $\kappa^{-1} \log (k^+)$. 
Instead, for $\w^+\approx10$, it seems to asymptote to a constant roughness function that is independent of roughness height.
The roughness function for the larger spacings ($\w^+\gtrsim20$) all appear to be entirely within this asymptotic regime, with $\Delta U^+$ being approximately equal to a constant that only depends on $\w^+$.
However, consider fixed values of $k/\w$, shown by the dashed and the dotted grey lines for $k/\w=3$ and $k/\w=6$, respectively. This is more representative of a laboratory study in which a single geometry (with fixed $k/\w$) is studied at various flow speeds which leads to the roughness Reynolds number varying. By considering fixed $k/\w$, the data now appear to behave more like a $k$-type roughness in that the larger values of $k^+$ are tending towards the fully rough asymptote. This is despite  the previous results showing very little change in the flow with such large aspect ratios. It would seem that the offset $D$ in 
$\kappa^{-1}\log(k^+)+D$ is dependent on $k/\w$, with increasing $k/\w$ leading to reduced $D$.

The roughness function as a function of solidity, $\Lambda=k/(2\w)$ (figure \ref{fig:DU}\emph{b}), shows how the roughness function for a particular $\w^+$ tends towards a constant value for large solidity (large $k/\w$ ratios). This value depends on the fluid width $\w^+$, with increasing $\w^+$ leading to an increasing roughness function. 
It is important to emphasise that here we are varying $k^+$ for a variety of $\w^+$ values. Previous studies that examine the roughness function for varying solidity typically do so by keeping $k^+$ fixed and varying $\w^+$. This results in the roughness function decreasing with solidity in the so-called dense regime when $\Lambda\gtrsim0.15$ \citep{Jimenez04,Flack14,MacDonald16}, as observed in figure \ref{fig:barSketch}(\emph{a}). If we were to consider a fixed $k^+$ value for the present bar roughness and increase the wavelength, we would still obtain a reducing roughness function in the dense regime ($\Lambda\gtrsim0.15$, shown by the vertical black dashed line in figure \ref{fig:DU}\emph{b}). To see this, consider the first data point for $\w^+\approx100$ (red cross, $k^+\approx50$, $\Lambda=0.25$) and final data point for $\w^+\approx10$ (black plus, $k^+\approx60$, $\Lambda=3$). Even though the roughness heights are slightly different, it is clear that the roughness function is reducing with solidity for these approximately matched $k^+$ values.

\setlength{\unitlength}{1cm}
\begin{figure}
\centering
	\includegraphics[width=0.46\textwidth]{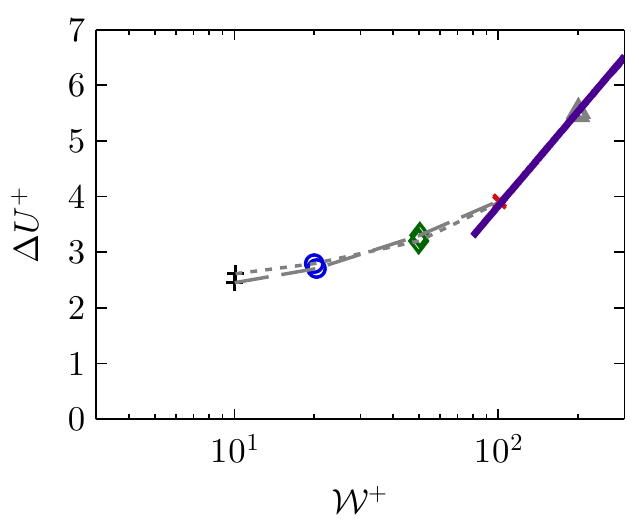}
	\includegraphics[width=0.49\textwidth]{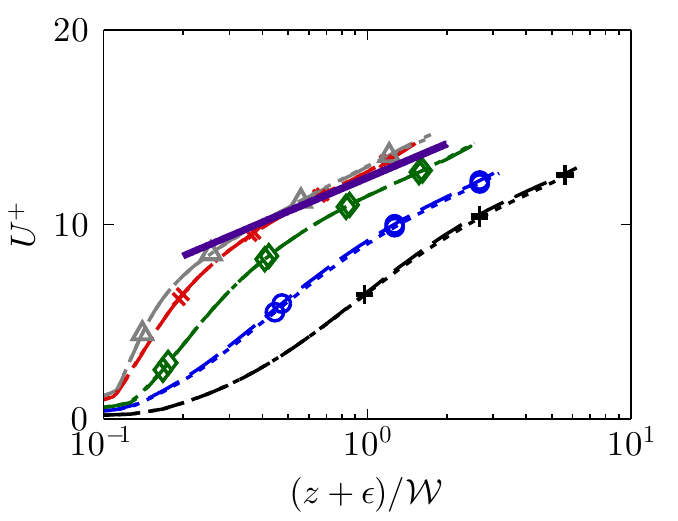}
	\put(-9.5,4.4){\small $\frac{1}{\kappa}\log(\w^+)-7.4$}
	\put(-5.5,4.0){\small $\frac{1}{\kappa}\log(z/\w)+12.4$}
	\put(-5.0,3.9){\line(1,-3){0.4}}
	\put(-1.7,3.8){Inc.~$\w^+$}
	\put(-0.9,3.5){\vector(-4,1){1.2}}
	\put(-13.2,4.75){(\emph{a})}
	\put(-6.6,4.75){(\emph{b})}
	\vspace{-0.5\baselineskip}
\caption{(Colour online) (\emph{a}) Roughness function $\Delta U^+$ against fluid width $\w^+$,  and 
(\emph{b}) mean velocity profiles against vertical position normalised on $\w^+$,
for solidity values 
\protect\raisebox{0.8ex}{\linethickness{0.5mm}\line(1,0){0.3}\hspace{0.15cm}\line(1,0){0.3}}, $\Lambda=1.5$
and
\protect\raisebox{0.8ex}{\linethickness{0.5mm}\line(1,0){0.1}\hspace{0.15cm}\line(1,0){0.1}\hspace{0.15cm}\line(1,0){0.1}}, $\Lambda = 3.0$ (figure \ref{fig:DU}\emph{b}).
Data only shown for the region of healthy turbulence, $z<z_c$.
Symbols:
\protect\raisebox{0.4ex}{\protect\scalebox{1.0}{$\color{black}\boldsymbol{\pmb{+}}$}}, $\w^+=10$;
\protect\raisebox{0.15ex}{\protect\scalebox{1.3}{$\color{myblue}\boldsymbol{\pmb{\circ}}$}}, $\w^+=20$;
\protect\raisebox{0.3ex}{\protect\scalebox{1.0}{$\color{mygreen}\boldsymbol{\pmb{\diamondsuit}}$}}, $\w^+=50$;
\protect\raisebox{0.3ex}{\protect\scalebox{1.0}{$\color{myred}\boldsymbol{\pmb{\times}}$}}, $\w^+=100$;
\protect\raisebox{0.1ex}{\protect\scalebox{1.0}{$\color{LGrey}\boldsymbol{\pmb{\bigtriangleup}}$}}, $\w^+=200$.
Solid purple line shows fully rough asymptote, $\Delta U^+=\kappa^{-1}\log({\w^+})+C$ in (\emph{a}) and $U^+=\kappa^{-1}\log(z/\w)+B_s-C$ in (\emph{b}), with $B_s\approx5.0$ and $C\approx-7.4$.
}
\label{fig:DUw}
\end{figure}

To better show the trend with $\w^+$, the roughness function is considered in figure \ref{fig:DUw}(\emph{a}) for surfaces with fixed aspect ratios of $k/\w=3$ and $k/\w=6$ (indicated by the vertical dashed and dotted lines in figure \ref{fig:DU}\emph{a}). For these fixed aspect ratio bars, the roughness function is seen to collapse. This suggests that for these high aspect ratio (tall and narrow) rough surfaces, it is $\w$, and not $k$ that is the relevant length scale. The $k$-type asymptote, $\kappa^{-1}\log(k^+)+B$ for fixed $k/\w$ would then become $\kappa^{-1}\log(\w^+)+C$, where $C=B+\kappa^{-1}\log(k/\w)$. If we assume the roughness function for $\w^+=200$ is in the fully rough regime, then this offset $C$ can be estimated as $C\approx -7.4$ from the present data and is independent of $k/\w$ (solid line in figure \ref{fig:DUw}). 
We can also evaluate
the equivalent sand grain roughness, $k/k_s \equiv\exp(-\kappa(3.5+B))=(k/\w)\exp(-\kappa(3.5+C))$.
With $C\approx -7.4$ for $k/\w\ge3$ we have 
$k_s \approx \w \exp(-3.9\kappa)\approx0.21\w$, 
indicating that $k_s$
is solely a function of $\w$.  The ratio $k_s/k$ can therefore be given as $k_s/k=0.11\Lambda^{-1}$, where $\Lambda=k/2\w$ is the solidity. This implies that for any bar roughness in this extremely dense regime of $\Lambda\ge1.5$, the equivalent sand grain roughness scales with $\Lambda^{-1}$.

Conventional $k$-type roughness in the fully rough regime exhibits a collapse in the mean velocity profile when plotted against the vertical position  normalised on roughness height, $k$. For the present roughness, the analogous scaling is to use the vertical position normalised on the bar spacing, $\w$. This is shown in figure \ref{fig:DUw}(\emph{b}) for both $k/\w=3$ and $k/\w=6$. Here, data is only shown for the region of healthy turbulence, $z<z_c=0.4L_y$, as the spanwise width of the minimal domain varies across different $\w^+$ cases.
In the near-wall region, there is a slight difference between the two cases with the largest spacings ($\w^+=100$ and $\w^+=200$), although this is likely because the flow has not quite reached the asymptotic fully rough state. Moreover, the logarithmic scaling of this plot exaggerates small differences close to the wall.
 However, there is a clear collapse in the logarithmic layer for $\w^+=100$ and $\w^+=200$, where the fully rough asymptote in mean velocity in the logarithmic layer is $U^+=\kappa^{-1}\log(z/\w)+B_s-C\approx\kappa^{-1}\log(z/\w)+12.4$, where $B_s\approx5.0$ is the smooth-wall log-law constant. This is simply an alternate representation of figure \ref{fig:DUw}(\emph{a}) and shows that the flow is becoming fully rough, where the dominant roughness length scale is the spacing between the bars, $\w$.

The results of figure \ref{fig:DUw} suggest that the flow has reached the asymptotic fully rough regime. However,
as discussed by \cite{Busse17}, even though bulk flow properties like $\Delta U^+$ and $U_b^+$ may suggest we have obtained this universal regime,
other flow properties such as the non-zero viscous drag (figure \ref{fig:FPratio}) and near-wall flow may still exhibit a Reynolds number dependence. We must 
therefore be careful and instead note that the flow appears to be tending towards this fully rough state, although it has not yet been comprehensively obtained.

\setlength{\unitlength}{1cm}
\begin{figure}
\centering
	\includegraphics[scale=0.95]{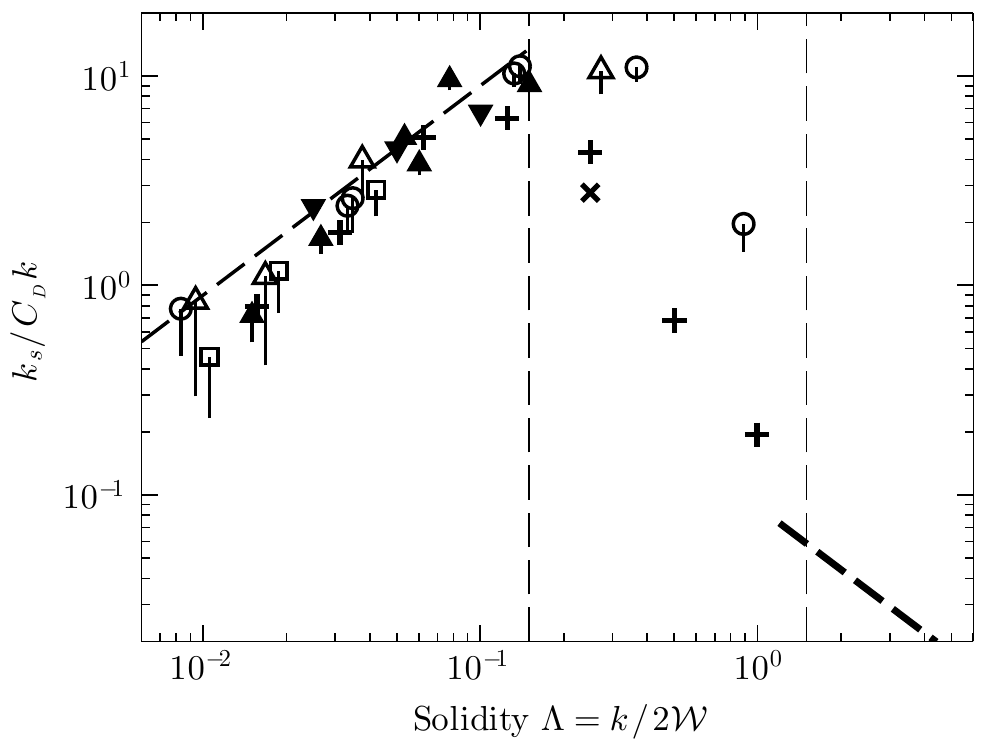}
	\put(-6.75,2.1){Sparse}
	\put(-3.6,2.1){Dense}
	\put(-1.3,2.25){Very}
	\put(-1.3,1.9){Dense}	
	\vspace{-0.5\baselineskip}
\caption{Equivalent sand grain roughness of several roughness geometries, adapted from figure 1 of \cite{Jimenez04}. The drag coefficient $C_D$ is to normalise for different roughness geometries, here we use $C_D\approx1.25$ as in \cite{Jimenez04}. The fully rough asymptote determined for the present data (figure \ref{fig:DUw}) is shown by the thick dashed line for $\Lambda\ge1.5$, where $k_s/k\approx0.11\Lambda^{-1}$.
}
\label{fig:jimWtype}
\end{figure}

Figure \ref{fig:jimWtype} shows the equivalent sand grain roughness against solidity for several different roughness geometries, adapted from figure 1 of \cite{Jimenez04}. The sparse regime ($\Lambda\le0.15$), which has been extensively studied, is seen to scale with $\Lambda$. Most authors term any roughness with $\Lambda>0.15$ to be dense, however no qualitative scaling for this regime has been proposed, and few experimental studies have been able to examine this regime. The present study has demonstrated that the equivalent sand grain roughness for spanwise-aligned rectangular bars scales with $\Lambda^{-1}$ for very dense rough surfaces with $\Lambda\gtrsim1.5$ (see figure \ref{fig:DUw}(\emph{a}) and associated discussion). This is shown in figure \ref{fig:jimWtype} by the thick dashed line, where we have used a drag coefficient of $C_D=1.25$, as recommended in \cite{Jimenez04} for spanwise bar roughness.
Note that if we define the drag coefficient with the slip velocity at the roughness crests, $C_D=\tau_w/\tfrac{1}{2}\rho U_k^2=2/{U_k^{+2}}$, then from figure \ref{fig:slipLength} we have $C_D\approx0.9$--1.3. This minor variation in $C_D$ only shifts the thick black line in figure \ref{fig:jimWtype} up or down slightly and does not affect the main result that $k_s/k$ is scaling with $\Lambda^{-1}$.

This leads to a new interpretation of figure \ref{fig:jimWtype}, where the sparse ($\Lambda\le0.15$) and very dense ($\Lambda\gtrsim1.5$) regions are separated by the dense region. In the well-studied sparse regime, the equivalent sand grain roughness increases with $\Lambda$, and the roughness height $k$ is the dominant roughness length scale. In the very dense regime, the equivalent sand grain roughness scales with $\Lambda^{-1}$ and we have shown that the spacing between the bars $\w$ is an important roughness length scale. This suggests that in the intermediary dense region ($0.15\le\Lambda\lesssim1.5$) both $k$ and $\w$ are competing roughness length scales, which may help in understanding why no scaling for $\Lambda\gtrsim0.15$ has received widespread support. It is likely that this dense regime is highly dependent on the roughness geometry. Moreover, the ratio of the bar width to wavelength, $b/\lambda$ is an important parameter, which here has been kept constant at $b/\w=1$ in the present study.

\subsection{Near-wall flow}
\setlength{\unitlength}{1cm}
\begin{figure}
\centering

	\includegraphics[width=0.49\textwidth]{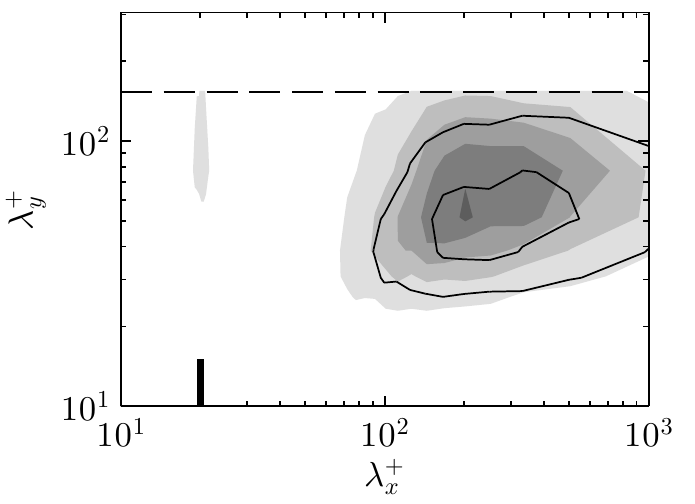}
	\includegraphics[width=0.49\textwidth]{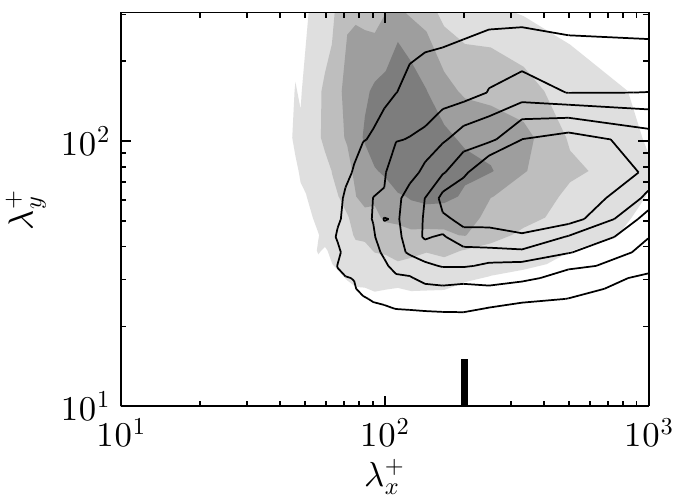}
	\put(-13.5,4.7){(\emph{a})}
	\put(-6.7,4.7){(\emph{b})}
	\put(-11.3,1.1){$\lambda^+$}
	\put(-2.05,1.1){$\lambda^+$}
	\vspace{-0.5\baselineskip}
\caption{Two-dimensional pre-multiplied energy spectrum of vertical (wall-normal) velocity for $k/\w=6$ (shaded contour) for (\emph{a}) $\w^+=10$ at $z^++\epsilon^+\approx6.4$ and (\emph{b}) $\w^+=100$ at $z^++\epsilon^+\approx 16.1$. This corresponds to a location approximately 5 viscous units above the roughness crests. The thick vertical line at the bottom of the plots marks the bar wavelength $\lambda^+=b^++\w^+=2\w^+$. Line contour shows smooth wall spectrum at matched $z^++\epsilon^+$ locations. Contour spacing is (\emph{a}) $0.002U_\tau^2$ and (\emph{b}) $0.005U_\tau^2$.
}
\label{fig:wallNormSpect}
\end{figure}

In order to analyse the changes to the flow structures that occur as we tend towards the fully rough regime of this extremely dense roughness, figure \ref{fig:wallNormSpect} shows the two-dimensional pre-multiplied energy spectra of vertical (wall-normal) velocity, $k_x k_yE_{w'w'}$. This is done for two cases with $k/\w=6$, at a horizontal (wall-parallel) location approximately 5 viscous units above the roughness crest. This spectrum is related to the vertical root-mean-square velocity fluctuations as $w_{rms}'^2=\int_0^\infty\int_0^\infty k_x k_y E_{w'w'}\id \log(k_x) \id \log(k_y)$, where single prime denotes purely turbulent fluctuations, with the temporally independent component $\widetilde{w}$ subtracted off. The case with the narrower bars ($\w^+=10$, figure \ref{fig:wallNormSpect}\emph{a}) has moderate agreement with the smooth-wall flow, in that the peak is located close to $\lambda_x^+\approx250$ and $\lambda_y^+\approx50$. This corresponds to the quasi-streamwise vortices \citep{Jeong97} that accompany the familiar near-wall streaks \citep{Kline67}. The strength of the energy spectrum in the rough-wall flow is much greater, as the roughness enables the vertical velocity to penetrate into the roughness canopy while the smooth-wall flow has the impermeability condition at $z^+=0$. 
However, when the roughness spacing is increased to $\w^+=100$ (figure \ref{fig:wallNormSpect}\emph{b}), we see that there is significant energy located at long spanwise wavelengths at $\lambda_x^+\approx100$ that is absent in the smooth-wall flow. This resembles the identification of long spanwise rollers over streamwise-aligned riblets \citep{Mayoral11jfm}, that leads to the degradation of the drag-reducing regime. These rollers develop due to an inflectional point in the mean velocity profile, which leads to a two-dimensional Kelvin--Helmholtz-like instability. These rollers have long been associated with canopy flows over high aspect ratio roughness \citep[e.g.][]{Ghisalberti04,Nepf12}, where \cite{Raupach96} demonstrated a similarity between the canopy shear flow and mixing layers.

\setlength{\unitlength}{1cm}
\begin{figure}
\centering

	\includegraphics[trim = 0 17 0 0,clip = true,width=0.49\textwidth]{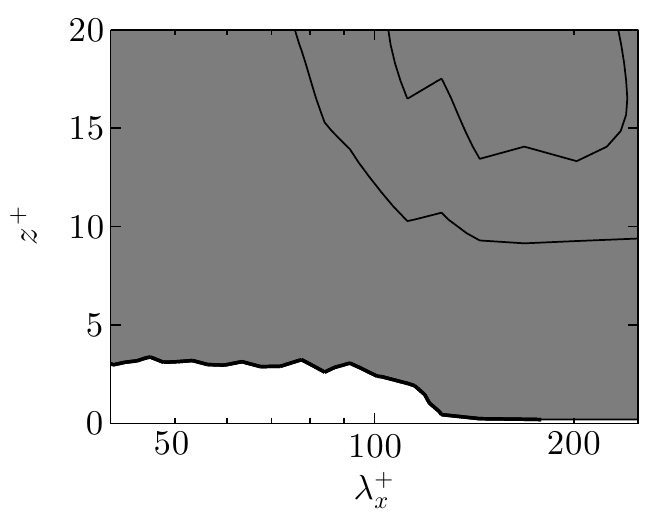}
	\includegraphics[trim = 0 17 0 0,clip = true,width=0.49\textwidth]{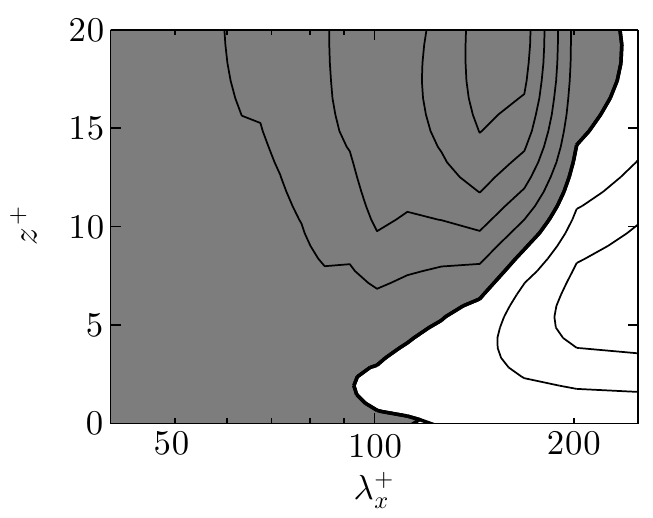} 
	\put(-13.4,4.3){(\emph{a})}
	\put(-6.6,4.3){(\emph{b})}
	\\
	\includegraphics[width=0.49\textwidth]{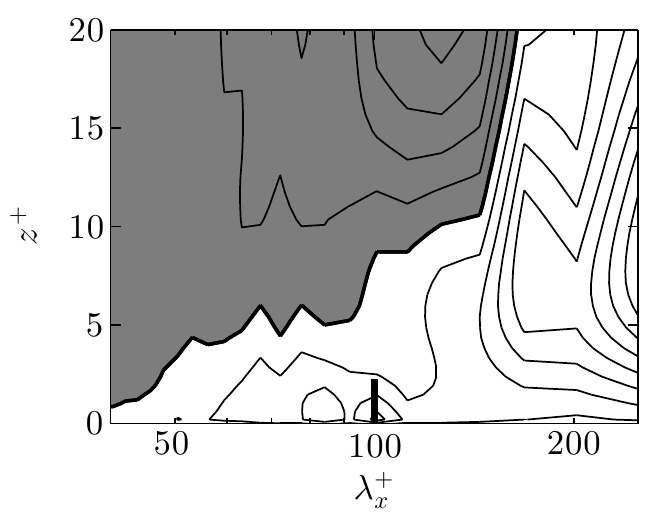}
	\includegraphics[width=0.49\textwidth]{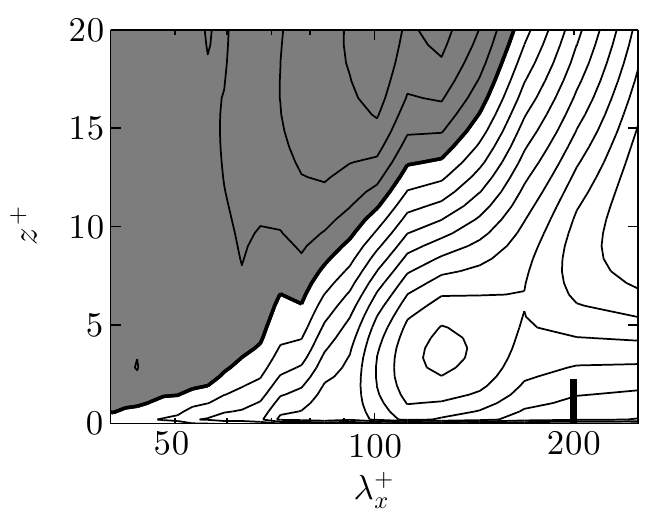}
	\put(-13.4,4.9){(\emph{c})}
	\put(-6.6,4.9){(\emph{d})}
	 \\	
	\vspace{-0.5\baselineskip}
\caption{Pre-multiplied streamwise cospectra of the Reynolds stress, $k_x E_{uw}^+$, where the cospectra have been integrated over $\lambda_y^+\ge 153$, for
(\emph{a}) smooth wall,
(\emph{b}) $\w^+=10$,
(\emph{c}) $\w^+=50$
and
(\emph{d})  $\w^+=100$.
All bar roughness cases are for $k/\w=6$.
The thick vertical line marks the bar wavelength $\lambda^+=2\w^+$
Contour spacing is $3\times 10^{-3}$, with the shaded area denoting positive values.
}
\label{fig:EuwIntLong}
\end{figure}

To better identify these long spanwise rollers, figure \ref{fig:EuwIntLong} shows the cospectra of the Reynolds stress, where the spectra have been integrated only over $\lambda_y^+\ge153$. This is similar to figure 11 of \cite{Mayoral11jfm}. The smooth-wall flow (figure \ref{fig:EuwIntLong}\emph{a}) is mostly positive (counter-gradient), however as the spacing between the bars increases the Reynolds stress becomes negative. It forms a local maxima in which the shear stress is concentrated around $z^+\approx 4$, $\lambda_x^+\approx130$ for $\w^+=100$ (figure \ref{fig:EuwIntLong}\emph{d}). Although this agrees well with the streamwise-aligned bars of \cite{Mayoral11jfm} in the drag-increasing regime, where the maxima was at $z^+\approx4$, $\lambda_x^+\approx150$, the present structures could originate from different mechanisms as these structures have a streamwise wavelength that is of the same order as the wavelength of the roughness. In any case, the existence of spanwise coherence in these spectrograms suggests that a  Kelvin-Helmholtz instability drives the shear layer that forms between each spanwise-aligned bar of the present roughness. 

\setlength{\unitlength}{1cm}
\begin{figure}
\centering
	\includegraphics[width=0.99\textwidth,trim = 0 20 0 0,clip=true]{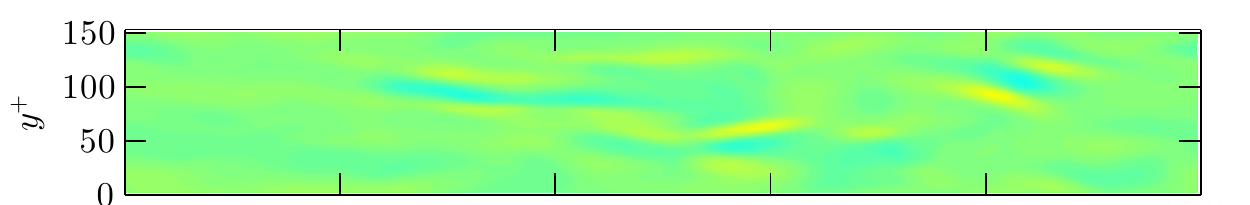}
	\put(-13.5,1.9){(\emph{a})}
	\\
	\includegraphics[width=0.99\textwidth,trim = 0 20 0 0,clip=true]{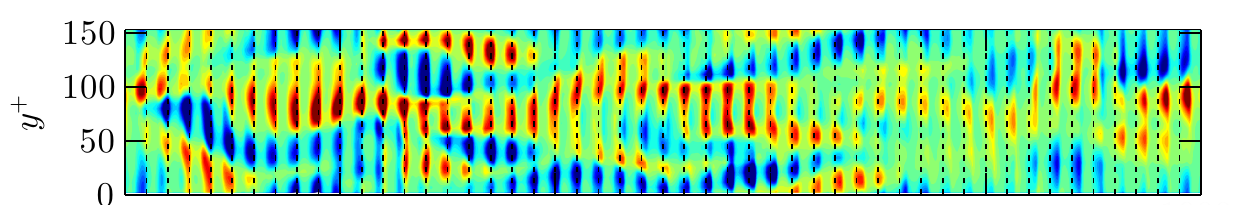}
	\put(-13.5,1.9){(\emph{b})}
	\\
	\includegraphics[width=0.99\textwidth,trim = 0 20 0 0,clip=true]{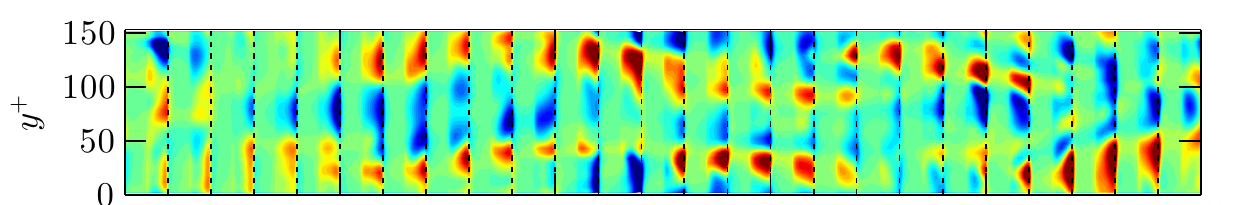}
	\put(-13.5,1.9){(\emph{c})}
	\\
	\includegraphics[width=0.99\textwidth,trim = 0 20 0 0,clip=true]{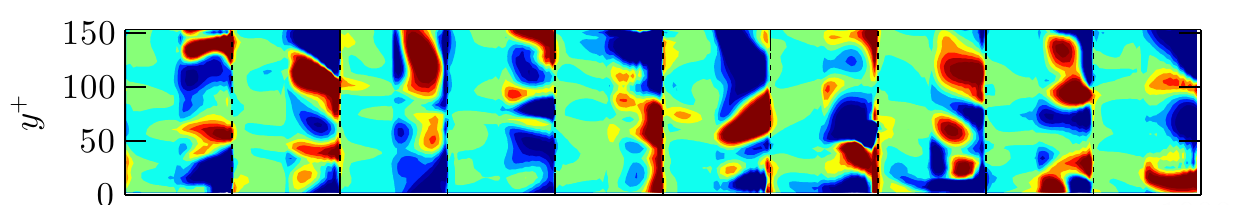}
	\put(-13.5,1.9){(\emph{d})}
	\\
	\includegraphics[width=0.99\textwidth,trim = 0 20 0 0,clip=true]{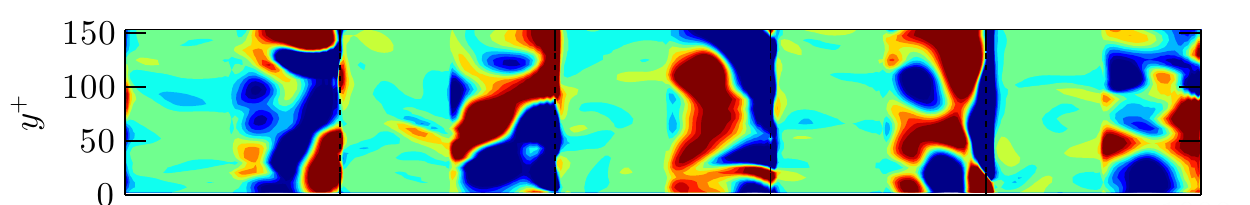}
	\put(-13.5,1.9){(\emph{e})}
	\\
	\includegraphics[width=1.0\textwidth,trim = 0 0 0 0,clip=true]{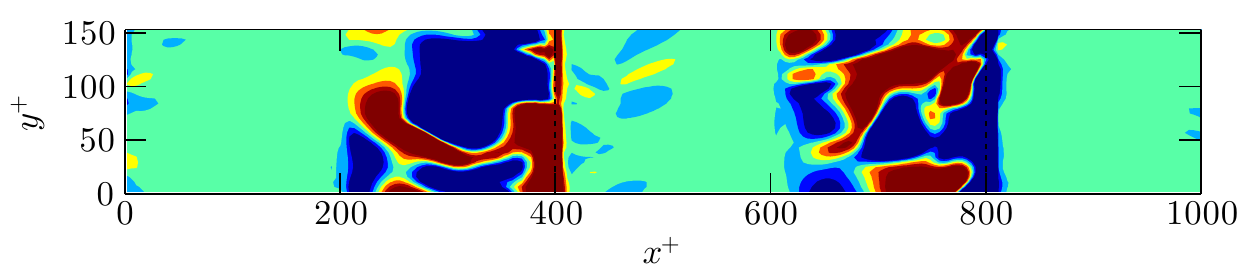}
	\put(-13.5,2.6){(\emph{f})}
	\\
	\includegraphics[trim = 0 0 0 0,clip=true]{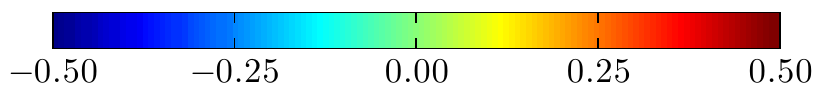}
	\put(-4.4,-0.3){$w'^+$}
	\vspace{0.2\baselineskip}
\caption{(Colour online). Contours of instantaneous vertical (wall-normal) turbulent fluctuations (with the time-independent dispersive component $\widetilde{w}$ removed) in a horizontal (wall-parallel) plane at $z^+\approx 2$, for $k/\w=6$ and
(\emph{a}) smooth wall,
(\emph{b}) $\w^+=10$,
(\emph{c}) $\w^+=20$,
(\emph{d}) $\w^+=50$,
(\emph{e}) $\w^+=100$
and
(\emph{f}) $\w^+=200$.
Vertical dotted lines show the leading edge of the bars.
}
\label{fig:vInst}
\end{figure}

An instantaneous snapshot of vertical turbulent fluctuations  (with the time-independent dispersive component $\widetilde{w}$ removed) at $z^+\approx 2$ is shown in figure \ref{fig:vInst}, for $k/\w=6$ bars. The smooth-wall flow (figure \ref{fig:vInst}\emph{a}) has near zero vertical velocity due to the impermeability constraint, although structures that are elongated in the streamwise direction can faintly be seen. The two narrower bars ($\w^+=10$ and $\w^+=20$, \emph{a} and \emph{b}) more clearly show this streaky structure. There is a definite coherence across the spanwise bars, even though there are strong fluctuations occurring on the leading edges of the bars. For $\w^+\ge50$, little coherence between adjacent bars remains to be seen at this scale, with the strong vertical velocity saturating across much of the fluid region between the bars. The above analyses suggests that the near-wall streaky structure has been replaced by spanwise rollers with $\lambda_x^+\approx150$, although it is difficult to observe this in instantaneous snapshots and with such a narrow spanwise domain.

%
%
\section{Conclusions}
\label{sect:discConc}

Direct numerical simulations of turbulent flow over spanwise-aligned bars have been performed in which the height of the bars, $k$, is larger than the spacing between them, $\w$. These bars are sometimes referred to as $d$-type roughness, as the roughness function is thought to scale on the outer-layer length scale. The present simulations are conducted in a minimal-span channel, which explicitly limits the size of the largest length scale in the flow. By progressively widening the channel and increasing this largest length scale, it was found that there was little change to the mean velocity profile when $L_y\ge3\w$, or when the critical height $z_c=0.4L_y\gtrsim1.2\w$. This suggests that the outer layer of the flow is not significant to this rough surface, raising questions about the classification of such surfaces as $d$-type.

The roughness function appears to be tending towards the fully rough $k$-type asymptote when fixed ratios of $k/\w$ are considered (figure \ref{fig:DU}\emph{a}), where each ratio of $k/\w$ has a different offset constant in $\kappa^{-1}\log(k^+)+B$. However, for fixed $\w^+$ the pressure to total drag ratio and virtual origin all show little variation with $k$, suggesting that $k$ is no longer relevant to the flow for these very deep bars. A  clear collapse is seen when the roughness function is instead plotted as a function of $\w^+$ for fixed ratios of $k/\w\ge3$. The fully rough asymptote would then be $\kappa^{-1}\log(\w^+)+C$, where the offset constant is estimated as $C\approx -7.4$.
For high aspect ratio bars with $k/\w\gtrsim3$,  the equivalent sand grain roughness can be estimated as $k_s = 0.21\w$. This applies for any bar roughness which has a sufficiently high aspect ratio $k/\w\ge3$. This suggests that these extremely dense bars lead to a $k_s/k\propto\Lambda^{-1}$ scaling where $\w$ is the dominant roughness length scale parameter for $\Lambda\gtrsim1.5$. It is proposed that the dense region ($0.15\le\Lambda\lesssim1.5$) between sparse and very  dense roughness is therefore the result of competing effects of the roughness height, $k$, and spacing between the roughness, $\w$. It is also likely that the start of the extremely dense regime is roughness geometry dependent, where the weaker sheltering of three-dimensional roughness would require a higher aspect ratio to reach this asymptotic regime.

\section*{Acknowledgements}
The authors would like to gratefully acknowledge the financial support of the Australian Research Council and the Bushfire and Natural Hazards Cooperative Research Council. MM was supported through an Australian Government Research Training Program Scholarship. Computational time was granted under the Victoria Life Sciences Computational Initiative, which is supported by the Victorian Government, Australia and the Pawsey Supercomputing Centre with funding from the Australian Government and the Government of Western Australia.


\bibliographystyle{jfm}
\bibliography{MMbib}

\end{document}